\documentclass[aps,prd,preprintnumbers,groupedaddress,nofootinbib,amssymb,eqsecnum,notitlepage]{revtex4}
\usepackage{graphicx}
\usepackage{bm}
\usepackage{amsmath}
\usepackage{color}
\usepackage{amsfonts}

\usepackage{graphicx}
\usepackage{bm}
\usepackage{amsmath}
\usepackage{color}
\usepackage{amsfonts}

\newcommand{\newc}{\newcommand}

\newcommand{\ben}{\begin{eqnarray}}
\newcommand{\een}{\end{eqnarray}}

\newc{\be}{\begin{equation}}
\newc{\ee}{\end{equation}}
\newc{\ba}{\begin{eqnarray}}
\newc{\ea}{\end{eqnarray}}
\newc{\bea}{\begin{eqnarray*}}
\newc{\eea}{\end{eqnarray*}}
\newc{\ie}{{\it i.e.} }
\newc{\eg}{{\it e.g.} }
\newc{\etc}{{\it etc.} }
\newc{\etal}{{\it et al.}}

\newc{\ra}{\rightarrow}
\newc{\lra}{\leftrightarrow}
\newc{\lsim}{\buildrel{<}\over{\sim}}
\newc{\gsim}{\buildrel{>}\over{\sim}}
\newc{\C}{{\cal C}}
\newc{\D}{{\cal D}}
\newc{\Xp}{X_{\phi}}
\newc{\Xc}{X_{\chi}}
\newc{\Mpl}{M_{\rm pl}}

\begin{document}
\preprint{YITP-16-36} 
\preprint{IPMU16-0032} 

\title{Cosmology in generalized Proca theories}

\author{Antonio De Felice$^{1}$,
Lavinia Heisenberg$^{2}$,
Ryotaro Kase$^{3}$,
Shinji Mukohyama$^{1,4}$,\\
Shinji Tsujikawa$^{3}$ and
Ying-li Zhang$^{5,6}$}

\affiliation{$^1$Yukawa Institute for Theoretical Physics, Kyoto University, 606-8502, Kyoto, Japan\\
$^2$Institute for Theoretical Studies, ETH Zurich, Clausiusstrasse 47, 8092 Zurich, Switzerland\\
$^3$Department of Physics, Faculty of Science, Tokyo University of Science, 1-3, Kagurazaka,
Shinjuku-ku, Tokyo 162-8601, Japan\\
$^4$Kavli Institute for the Physics and Mathematics of the Universe (WPI), 
The University of Tokyo Institutes for Advanced Study, The University of Tokyo, Kashiwa, Chiba 277-8583, Japan\\ 
$^5$National Astronomy Observatories, Chinese Academy of Science,
Beijing 100012, People's Republic of China\\
$^6$Institute of Cosmology and Gravitation,
University of Portsmouth, Portsmouth PO1 3FX, UK}

\date{\today}

\begin{abstract}

We consider a massive vector field with derivative interactions that propagates
only the 3 desired polarizations
(besides two tensor polarizations from gravity) 
with second-order equations
of motion in curved space-time. The cosmological implications of such
generalized Proca theories are investigated for both the background and
the linear perturbation by taking into account the
Lagrangian up to quintic order.
In the presence of a matter fluid with a temporal component of 
the vector field, we derive the background equations of motion and show the 
existence of de Sitter solutions relevant to the late-time cosmic acceleration. 
We also obtain conditions for the
absence of ghosts and Laplacian instabilities of tensor, vector, and
scalar perturbations in the small-scale limit.
Our results are applied to concrete examples of the general functions
in the theory, which encompass vector Galileons as a specific case.
In such examples, we show that the de Sitter fixed point is always
a stable attractor and study viable parameter spaces in which
the no-ghost and stability conditions are satisfied during the
cosmic expansion history.

\end{abstract}

\pacs{04.50.Kd,95.30.Sf,98.80.-k}

\maketitle

\section{Introduction}

The high-precision observations achieved by the measurements like supernovae Type Ia (SNIa) \cite{SNIa},
Cosmic Microwave Background (CMB) \cite{CMB}, and Baryon Acoustic Oscillations (BAO) \cite{BAO}
together with the two pillars of General Relativity (GR) and the cosmological principle
have led to a notion of the standard cosmological model.  According to this concordance model,
about 70\,\% and 25\,\% of the today's energy density of the Universe
correspond to shadowy components dubbed dark energy and
dark matter, respectively.
In particular, dark energy has a repulsive force against gravity,
which leads to the late-time cosmic acceleration.
After its first discovery in 1998, there have been numerous attempts to pursue the origin of dark energy \cite{CST}.

The first promising explanatory attempt for dark energy consists of introducing a cosmological constant $\Lambda$ that would cause an effective repulsive force between cosmological objects at large distances. A natural origin of the cosmological constant could be the vacuum energy density.
Using the technique of particle physics, one can estimate the
expected value for the vacuum energy density
caused by fluctuating quantum fields \cite{Weinberg}.
The result is the worst theoretical prediction in the history of physics. The discrepancy between the theoretically predicted value and the measured one amounts to 120 orders of magnitude. This constitutes the vacuum catastrophe prediction which remains one of the most challenging puzzles in physics.

An alternative explanation for the late-time acceleration of the Universe could be accounted by introducing new dynamical degrees of freedom. This is achieved either by directly invoking new fluids in form of an energy-momentum tensor
$T_{\mu\nu}$ with negative pressure or by modifying the geometrical part of Einstein field equations 
(see Refs.~\cite{moreview} for reviews).
One hope in weakening gravity on cosmological scales is to tackle the cosmological constant problem. The implications are multifaceted. The same modification might account for the late-time speed-up of the cosmic expansion.
This type of scenarios can naturally arise in higher-dimensional theories
or massive gravity.

Concerning the higher-dimensional frameworks, the
Dvali-Gabadadze-Porrati (DGP) model is one of the most important
large-scale modifications of gravity \cite{DGP}.
The Universe is modeled by a confined three-brane embedded in a five-dimensional Minkowski bulk. 
An intrinsic Einstein-Hilbert term is introduced to recover four-dimensional gravity on small scales.
On the other hand, the gravity is systematically weakened on cosmological scales due to the 
gravitational leakage to the extra dimension. From the effective four dimensional point of view, 
the graviton acquires a soft mass and hence carries five degrees of freedom. 
One of such degrees of freedom corresponds to  the helicity-0 mode, 
which sources the existence of a self-accelerating solution.

The non-linear interactions of the helicity-0 mode make the mode
decouple from the gravitational dynamics, which is known as the 
Vainshtein mechanism \cite{Vainshtein:1972sx}. 
Motivated by the nice properties of the helicity-0 mode in
the decoupling limit of the DGP model, more general
``Galileon'' interactions have been proposed
on the Minkowski background \cite{Nicolis:2008in}.
These interactions are invariant under internal
Galilean and shift transformations.
The Galilean symmetries together with the requirement of ghost absence restrict the effective Lagrangian to consist
only of five derivative interactions.

There were attempts to generalize the Minkowski Galileon theory to a curved background. The first natural pursuit was done by covariantizing directly the decoupling limit \cite{Chow:2009fm}.
A rigorous investigation revealed that the naive covariantization (replacing
partial derivatives with covariant derivatives) would yield
higher-order equations of motion and that unique non-minimal couplings with the curvature
should be added to maintain the equations of motion
up to second order \cite{Deffayet:2009wt}.
The generalization of this latter ``covariant Galileon'' led to the
rediscovery of the Horndeski action \cite{Horndeski} as the most
general scalar-tensor theories with second-order equations of motion
in $4$-dimensions \cite{Horndeski2}.
The application of covariant Galileons to cosmology
witnessed a flurry of investigations concerning self-accelerating de Sitter solutions \cite{Silva:2009km}, late-time cosmology and its observational
implications \cite{Sami,DT10}, inflation
\cite{Creminelli:2010ba,Burrage:2010cu,Mizuno:2010ag},
super-luminalities \cite{Hinterbichler:2009kq}, and so on.

Even if the ghost-like Ostrogradski instability \cite{Ostro} can be avoided for covariant Galileons, the Galilean symmetry is explicitly broken when promoted to the curved space-time. However, there have been also successful generalizations to maximally symmetric backgrounds with a generalized Galilean
symmetry \cite{Goon:2011qf}.
It is worth mentioning that there exists a parallel to massive gravity, namely Galileon interactions 
arise naturally in the decoupling limit of massive gravity \cite{deRham:2010ik}. 
Furthermore, the covariantization of the decoupling-limit interactions gives rise to 
a specific sub-class of Horndeski theories \cite{deRham:2011by}.

A natural follow-up was to apply the same construction of Galileon-like interactions 
to arbitrary $p$-forms \cite{Deffayet:2010zh}. 
While even $p$-form Galileons and multi odd $p$-form Galileons are easy to construct, 
very soon it was realized that one cannot construct Galilean
interactions for a Lorentz- and gauge-invariant, 
single spin-$1$ (i.e., $1$-form) field in any dimensions \cite{Gum} (see also
Refs.~\cite{Horndeski3,Barrow,Jimenez}). 
The formalism developed for the proof of the no-go theorem was recently 
extended towards categorization of all possible general odd $p$-form Galileons, 
as an application of the concept called plethysm in the representation theory of 
the symmetric groups \cite{Deffayet:2016von}.
However, this no-go theorem does not apply to massive spin 1 fields
since one of the assumptions of the theorem, the gauge invariance, is dropped, 
and it is possible to successfully construct derivative
self-interactions for the vector field with a broken $U(1)$ symmetry.
A systematical construction of these interactions with only 3 propagating
degrees of freedom (two transverse and one longitudinal)
was performed in Ref.~\cite{Heisenberg} with the requirement that the longitudinal mode possesses non-trivial interactions belonging to the class of Galileon/Horndeski theories. 
The analysis was based on studying the Hessian matrix and assuring 
the propagation of a second class constraint.

The action derived in Ref.~\cite{Heisenberg} corresponds to
the generalization of massive Proca theories to the curved background
in which the requirement of second-order equations of motion enforces 
the presence of non-minimal derivative couplings with
the Ricci scalar and the Einstein tensor \cite{Heisenberg,Tasinato}
(see also Refs.~\cite{TKK,Fleury,Hull,Li} for related works).
Note that similar type of vector-tensor theories can naturally arise
in modifications of gravity based on Weyl geometries \cite{Jimenez:2014rna}.
It is also possible to obtain derivative interactions higher than
quintic order by keeping the degrees of
freedom three \cite{Peter,Jimenez:2016isa}.

In a sub-class of generalized Proca theories up to the quartic Lagrangian
${\cal L}_4$, the existence of de Sitter solutions relevant to dark energy
was found in Refs.~\cite{Tasinato}.
Applying these theories to the spherically symmetric background, the
authors in Ref.~\cite{scvector} showed that cubic-order derivative
self-interactions and non-minimal derivative couplings at quartic order
can efficiently screen the propagation of the fifth force in local regions
of the Universe. In this paper, we will closely study cosmological implications
of the full derivative interactions introduced in Ref.~\cite{Heisenberg}
in the presence of matter (radiation and non-relativistic matter).
In particular, we derive conditions for the absence of ghosts
and Laplacian instabilities by considering tensor, vector, and scalar
perturbations at linear level.
Our general results will be applied to concrete
models relevant to the late-time cosmic acceleration.
Not only we show the existence of stable de Sitter solutions, but also we
study the stability of perturbations throughout the cosmological evolution.

Our paper is organized as follows.
In Sec.~\ref{HPsec} we present the background equations of motion
in the presence of a temporal component of the vector field and
a matter fluid.
In Sec.~\ref{spsec} we derive conditions for avoiding ghosts and
Laplacian instabilities by expanding the generalized Proca action up to
quadratic order in tensor, vector, and scalar perturbations.
In Sec.~\ref{genedissec} we discuss general conditions for the
theoretical consistency at the de Sitter fixed point
and in the early cosmological epoch.
In Sec.~\ref{covasec} we propose a class of concrete models in which the de Sitter
point is always a late-time attractor and discuss the viable parameter
space in which ghosts and instabilities are absent during the cosmic
expansion history.
Sec.~\ref{consec} is devoted to conclusions.

\section{Generalized Proca theories and the background
equations of motion}
\label{HPsec}

Let us start with generalized Proca theories described by
the four-dimensional action \cite{Heisenberg}
\be
S=\int d^4x \sqrt{-g} \left( {\cal L}
+{\cal L}_M \right)\,,\qquad
{\cal L}={\cal L}_F+\sum_{i=2}^{5} {\cal L}_i\,,
\label{Lag}
\ee
where $g$ is a determinant of the metric tensor
$g_{\mu \nu}$, ${\cal L}_M$ is a matter
Lagrangian, and
\ba
{\cal L}_F &=& -\frac14 F_{\mu \nu}F^{\mu \nu}\,,
\label{LF}\\
{\cal L}_2 &=& G_2(X)\,,
\label{L2}\\
{\cal L}_3 &=& G_3(X) \nabla_{\mu}A^{\mu}\,,
\label{L3}\\
{\cal L}_4 &=&
G_4(X)R+
G_{4,X}(X) \left[ (\nabla_{\mu} A^{\mu})^2
+c_2 \nabla_{\rho}A_{\sigma} \nabla^{\rho}A^{\sigma}
-(1+c_2) \nabla_{\rho}A_{\sigma}
\nabla^{\sigma}A^{\rho} \right]\,,\label{L4} \\
{\cal L}_5 &=&
G_{5}(X) G_{\mu \nu} \nabla^{\mu} A^{\nu}
-\frac16 G_{5,X}(X) [ (\nabla_{\mu} A^{\mu})^3
-3d_2 \nabla_{\mu} A^{\mu}
\nabla_{\rho}A_{\sigma} \nabla^{\rho}A^{\sigma}
-3(1-d_2) \nabla_{\mu} A^{\mu}
\nabla_{\rho}A_{\sigma} \nabla^{\sigma}A^{\rho}
\nonumber \\
& &
+(2-3d_2) \nabla_{\rho}A_{\sigma} \nabla^{\gamma}
A^{\rho} \nabla^{\sigma}A_{\gamma}
+3d_2 \nabla_{\rho}A_{\sigma} \nabla^{\gamma}
A^{\rho} \nabla_{\gamma}A^{\sigma}]\,.
\label{L5}
\ea
Here, $A_\mu$ is a vector field with $F_{\mu \nu}=
\nabla_{\mu}A_{\nu}-\nabla_{\nu}A_{\mu}$,
$\nabla_{\mu}$ is the covariant derivative operator,
$R$ is the Ricci scalar, $G_{\mu \nu}$ is the
Einstein tensor, $c_2, d_2$ are constants,
$G_{2,3,4,5}$ are arbitrary functions of
\be
X=-\frac12 A_{\mu} A^{\mu}\,,
\label{Xdef}
\ee
and $G_{i,X}=\partial G_{i}/\partial X$.
The Lagrangians ${\cal L}_{2,3,4,5}$ are constructed to keep
the propagating degrees of freedom up to three with
second-order equations of motion.
As in standard massless Maxwell theory, there are two transverse
polarizations for the vector field. The Proca theory corresponds to
the function $G_2(X)=m^2 X$ with $G_{3,4,5}=0$, in which
case introduction of the mass term $m$ breaks the $U(1)$
gauge symmetry.
This gives rise to one additional degree of freedom
in the longitudinal direction.

The generalized Proca theories given above correspond to
the extension of massive Proca theories with the cubic
derivative self-interaction ${\cal L}_3$ and non-minimal
derivative couplings with the Ricci scalar (in the Lagrangian
${\cal L}_4$) and with the Einstein tensor (in the Lagrangian
${\cal L}_5$). In Eqs.~(\ref{L4}) and (\ref{L5}) the terms
multiplied by the constants $c_2$ and $d_2$ can be expressed
in terms of $F_{\mu \nu}$ \cite{Heisenberg,Peter},
so they can be regarded as intrinsic
vector modes (i.e., they vanish by taking the scalar limit
$A^{\mu} \to \nabla^{\mu}\pi$).
In Refs.~\cite{Tasinato} the cosmology for $c_2=-1$ up to
the Lagrangian ${\cal L}_4$ was studied for specific choices
of the functions $G_{2,3,4}$.
In this paper we study the cosmology for the full action
(\ref{Lag}) for arbitrary constants $c_2$ and $d_2$.
We also derive the background equations of motion and
the no-ghost/stability conditions without specifying
the functional forms of $G_{2,3,4,5}$.

We define the energy-momentum tensor of the matter
Lagrangian ${\cal L}_M$, as
\be
T_{\mu \nu}^{(M)}=-\frac{2}{\sqrt{-g}}
\frac{\delta(\sqrt{-g}{\cal L}_M)}
{\delta g^{\mu \nu}}\,.
\ee
Assuming that matter is minimally coupled to gravity,
the following continuity equation holds
\be
\nabla^{\mu}T_{\mu \nu}^{(M)}=0\,.
\label{Tcon}
\ee

We derive the equations of motion on the flat
Friedmann-Lema\^{i}tre-Robertson-Walker (FLRW) background 
described by the line element 
$ds^2=-dt^2+a^2(t)\delta_{ij}dx^idx^j$, where $a(t)$
is the scale factor that depends on the cosmic time $t$.
The homogenous vector field $A^{\mu}$, which does not break
spatial isotropy, is given by
\be
A^{\mu}=(\phi(t),0,0,0)\,,
\ee
\label{Aansatz}
where the temporal component $\phi$ is a function of $t$.
Then, the kinetic term (\ref{Xdef}) reduces to $X=\phi^2/2$.

For the matter Lagrangian ${\cal L}_M$,  we take into
account the perfect fluid with the energy-momentum tensor
$T^{\mu}_{\nu}={\rm diag}(-\rho_M,P_M,P_M,P_M)$,
where $\rho_M$ is the energy density and $P_M$
is the pressure. Then, the continuity equation
(\ref{Tcon}) reads
\be
\dot{\rho}_M+3H(\rho_M+P_M)=0\,,
\label{continuity}
\ee
where a dot represents a derivative with respect to $t$, and
$H \equiv \dot{a}/a$ is the Hubble expansion rate.

Varying the action (\ref{Lag}) with respect to $g_{\mu \nu}$,
we obtain the following equations of motion
\ba
& &
G_2-G_{2,X}\phi^2-3G_{3,X}H \phi^3
+6G_4H^2-6(2G_{4,X}+G_{4,XX}\phi^2)H^2\phi^2
+G_{5,XX} H^3\phi^5+ 5G_{5,X} H^3\phi^3
=\rho_M\,,
\label{be1}\\
& &
G_2-\dot{\phi}\phi^2G_{3,X}+2G_4\,(3H^2+2\dot{H})
-2G_{4,X}\phi \, ( 3H^2\phi +2H\dot{\phi}
+2\dot{H} \phi )-4G_{4,XX}H\dot{\phi}\phi^3\nonumber\\
&& {}+G_{5,XX}H^2\dot{\phi} \phi^4+G_{5,X}
H \phi^2(2\dot{H}\phi +2H^2\phi+3H\dot{\phi})
=-P_M\,.
\label{be2}
\ea

Variation of the action (\ref{Lag}) with respect to
$A^{\mu}$ leads to
\be
\phi \left( G_{2,X}+3G_{3,X}H\phi +6G_{4,X}H^2
+6G_{4,XX}H^2\phi^2
-3G_{5,X}H^3\phi-G_{5,XX}H^3 \phi^3 \right)=0\,.
\label{be3}
\ee
Among four Eqs.~(\ref{continuity})-(\ref{be3}), three of
them are independent. We note that Eqs.~(\ref{be1}) and (\ref{be3})
do not contain any time derivatives of  $\phi$.
This reflects the fact that the theory given by the action (\ref{Lag})
has been constructed to avoid the appearance of the propagating
degrees of freedom more than three.

{}From Eq.~(\ref{be3}) there are two branches of solutions.
One of them is $\phi=0$, but in this case the temporal component
of the vector field does not contribute to the background dynamics.
Another branch corresponds to
\be
G_{2,X}+3G_{3,X}H\phi +6G_{4,X}H^2
+6G_{4,XX}H^2\phi^2
-3G_{5,X}H^3\phi-G_{5,XX}H^3 \phi^3=0\,,
\label{be3d}
\ee
in which case the field $\phi$ is directly related to
the Hubble parameter $H$.
This allows the existence of de-Sitter solutions characterized by
constant $H$ and $\phi$, so we shall focus on the second branch
satisfying Eq.~(\ref{be3d}) in the following discussion.

\section{Conditions for avoiding ghosts and instabilities}
\label{spsec}

We derive conditions for the absence of ghosts and instabilities
of tensor, vector, and scalar perturbations on the flat FLRW
background. We consider two scalar metric perturbations
$\alpha, \chi$ and one vector perturbation $V_i$
by choosing the so-called flat gauge.
Under this choice, the temporal and spatial components
of gauge transformation vectors are fixed.
Taking into account the tensor perturbation $h_{ij}$,
the linearly perturbed line-element is given by \cite{Bardeen,KS,Mukhanov}
\be
ds^{2}=-(1+2\alpha)\,dt^{2}+2\left( \partial_{i}\chi
+V_i \right)dt\,dx^{i}+a^{2}(t) \left( \delta_{ij}
+h_{ij} \right)dx^i dx^j\,,
\ee
where $V_i$ and $h_{ij}$ satisfy the following conditions
\ba
& & \partial^i V_i=0\,,
\label{traconve1}\\
& & \partial^i h_{ij}=0\,,\qquad {h_i}^i=0\,.
\label{tracon}
\ea

As for the Proca vector field $A^{\mu}$, the time component
$A^0$ and the spatial vector $A^{i}$ can be expressed
in the following forms
\ba
A^{0} & = & \phi(t)+\delta\phi\,,\\
A^{i} & = & \frac{1}{a^{2}} \delta^{ij} \left(
\partial_{j}\chi_{V}+E_j \right)\,,
\ea
where $\delta \phi$ is the perturbation in $A^0$ (which
depends on both $t$ and $x^i$).
The perturbation $\chi_V$ corresponds to the intrinsic
scalar part, whereas $E_j$ is the vector part satisfying the
transverse condition
\be
\partial^j E_j=0\,.
\label{traconve2}
\ee

For the matter action $S_M=\int d^4 x \sqrt{-g}\,{\cal L}_M$, 
we consider a single perfect fluid 
with the energy density $\rho_M$.
For the description of the perfect fluid, we make use of the
Schutz-Sorkin action \cite{Sorkin} (see also Ref.~\cite{DGS}):
\be
S_M=-\int d^{4}x \left[ \sqrt{-g}\,\rho_M(n)
+J^{\mu}(\partial_{\mu}\ell+\mathcal{A}_1
\partial_{\mu}\mathcal{B}_1+\mathcal{A}_2
\partial_{\mu}\mathcal{B}_2) \right]\,,
\label{Spf}
\ee
where $\ell$ is a scalar, $J^\mu$ is a vector field of
weight one, $\mathcal{A}_1, \mathcal{A}_2, \mathcal{B}_1, \mathcal{B}_2$
are scalars whose perturbations are meant to describe the vector modes, 
and $n$ is the number density of the fluid defined by
\be
n=\sqrt{\frac{J^{\alpha}J^{\beta}g_{\alpha\beta}}{g}}\,.
\ee
Due to the transverse condition for the vector modes,
a third component $\mathcal{A}_3\partial_{\mu}\mathcal{B}_3$
is redundant in the action (\ref{Spf}).

We express $\ell$ and $J^{\mu}$ into the background and
perturbed components, as \cite{DGS,DeFelice:2015moy}
\ba
\ell &=&-\int^{t} \rho_{M,n}\,dt'
- \rho_{M,n} \,v\,,\\
J^{0} &=&  \mathcal{N}_{0}+\delta J\,,\\
J^{i} &=& \frac{1}{a^{2}}\delta^{ik}
\left( \partial_{k}\delta j+W_k \right)\,,
\ea
where $ \rho_{M,n} \equiv \partial\rho_M/\partial n$ is evaluated 
on the background, $\mathcal{N}_{0}=n_0 a^{3}$ is a constant
associated with the total number of fluid particles
($n_0$ is the background value of $n$), and $v$, $\delta J$
are the perturbations of $\ell$, $J^0$, respectively.
On the FLRW background, the action (\ref{Spf}) reduces to 
$S_{M}^{(0)}=\int d^4 x \sqrt{-g}\,(n_0\rho_{M,n}-\rho_M)$.
Since the term inside the bracket corresponds to the pressure 
$P_M$ of the fluid, we have 
\be
P_M=n_0\rho_{M,n}-\rho_M\,.
\label{PM}
\ee

The perturbation $\delta j$ corresponds to the scalar
part of $J^i$, whereas the intrinsic vector perturbation
$W_k$ obeys
\be
\partial^k W_k=0\,.
\label{traconve3}
\ee
We write the perturbation of ${\cal A}_i$ in the form
\be
\delta\mathcal{A}_i=\rho_{M,n}\,u_i\,,
\label{delAi}
\ee
where $u_i$ corresponds to the intrinsic vector part of
the four velocity $u^{\alpha}=J^{\alpha}/(n\sqrt{-g})$,
satisfying
\be
\partial^i u_i=0\,.
\label{traconve4}
\ee

It should be pointed out that the above form of action is
not the only possibility for describing the perfect fluid,
e.g., the k-essence form \cite{kes} is also another possibility.
However, in the theories under consideration, the
Schutz-Sorkin action is convenient for properly
accommodating  vector perturbations.
Moreover, it provides a natural and convenient choice of variables 
for the dust-fluid perturbation with a vanishing propagation speed 
squared $c_m^2$. On the other hand, in the k-essence form, 
one may need to perform a change of variables corresponding
to a canonical transformation, e.g., from the perturbation of the scalar field
to the density contrast, before taking the $c_m^2\to 0$ 
limit \cite{DeFelice:2015moy}. 

Due to the decomposition into tensor, vector, and scalar modes
explained above, expansion of the action (\ref{Lag}) up to second order 
in perturbations leads to the quadratic action of the form 
\be
S^{(2)}=S_T^{(2)}+S_V^{(2)}+S_S^{(2)}\,,
\ee
where $S_T^{(2)},S_V^{(2)},S_S^{(2)}$ correspond to contributions 
from tensor, vector, and scalar perturbations, respectively,
Variations of the action $S_T^{(2)},S_V^{(2)},S_S^{(2)}$ with respect to 
perturbed quantities give rise to the linear perturbation equations of motion 
for tensor, vector, and scalar modes. The same equations of motion can be 
derived after obtaining the general field equations by varying the action 
(\ref{Lag}) with respect to $g^{\mu \nu}$ and then decomposing into 
tensor, vector, and scalar perturbations (as in the case of GR discussed 
in Ref.~\cite{Mukhanov}). 
The advantage of using the second-order action is that 
no-ghost and stability conditions for dynamical fields 
can be easily deduced after integrating out all non-dynamical 
fields from the action.
In the following, we shall separately derive 
$S_T^{(2)},S_V^{(2)},S_S^{(2)}$ to discuss conditions for 
the absence of ghosts and instabilities.

\subsection{Tensor perturbations}

The tensor perturbation $h_{ij}$, which satisfies the transverse and
traceless conditions (\ref{tracon}), can be expressed in terms of
the two polarization modes $h_{+}$ and $h_{\times}$, as
$h_{ij}=h_{+}e_{ij}^{+}+h_{\times} e_{ij}^{\times}$.
In Fourier space, the unit bases $e_{ij}^{+}$ and $e_{ij}^{\times}$ 
obey the relations $e_{ij}^{+}({\bm k}) e_{ij}^{+}(-{\bm k})^*=1$,
$e_{ij}^{\times}({\bm k}) e_{ij}^{\times}(-{\bm k})^*=1$,
and $e_{ij}^{+}({\bm k}) e_{ij}^{\times}(-{\bm k})^*=0$,
where  ${\bm k}$ is the comoving wavenumber.

Expanding Eq.~(\ref{Lag}) up to second order
in tensor perturbations, the resulting second-order
action is given by
\be
S_T^{(2)}=\sum_{\lambda={+},{\times}}\int dt\,d^3x\,
a^3\,\frac{q_T}{8}  \left[\dot{h}_\lambda^2
-\frac{c_T^2}{a^2}(\partial h_\lambda)^2\right]\,,
\ee
where
\be
q_T=2G_4-2\phi^{2}G_{{4,X}}+
H\phi^{3}G_{{5,X}}\,,
\label{qT}
\ee
and
\be
c_{T}^{2}=
\frac {2G_{4}+\phi^{2}\dot\phi\,G_{{5,X}}}{q_T}\,.
\label{cT}
\ee
Note that $c_{T}^2$ corresponds to the propagation
speed squared of the tensor mode.
The ghost is absent under the condition $q_T>0$.
The Laplacian instability on small scales can be avoided
for $c_T^2>0$. 

\subsection{Vector perturbations}
\label{vecsec}

The four quantities $V_i$, $E_i$, $W_i$, and $u_i$ obey the transverse
conditions (\ref{traconve1}), (\ref{traconve2}), (\ref{traconve3}),
and (\ref{traconve4}), respectively.  For the practical computations
of the second-order action, we can consider the vector fields
depending on $t$ and $z$ alone, e.g.,
$V_i=(V_{1}(t, z), V_{2}(t,z), 0)$.  Then, the transverse condition
such as Eq.~(\ref{traconve1}) is automatically satisfied.  For the
quantities ${\cal A}_i$ and ${\cal B}_i$, the simplest choice
containing all the needed information of the vector mode is given by
${\cal A}_1=\delta {\cal A}_1(t,z)$,
${\cal A}_2=\delta {\cal A}_2(t,z)$,
${\cal B}_1=x+\delta {\cal B}_1(t,z)$, and
${\cal B}_2=y+\delta {\cal B}_2(t,z)$ \cite{DGS}.  
The normalizations of ${\cal B}_1$ and ${\cal B}_2$ have been done such that 
both ${\cal B}_{1,x}$ and ${\cal B}_{2,y}$ are equivalent to 1.
Note that the above prescription can be extended to the general case in which the
perturbations depend on $t,x,y,z$.

After expanding the matter action (\ref{Spf}) up to second order in 
vector perturbations, it follows that 
\be
(S_V^{(2)})_M=\sum_{i=1}^{2}\int dtd^3x
\left[ \frac{1}{2a^2{\cal N}_0} \left\{
\rho_{M,n} \left (W_i^2+{\cal N}_0^2 V_i^2 \right)
+{\cal N}_0 \left(2\rho_{M,n}V_iW_i-a^3\rho_M V_i^2 \right) 
\right\}-{\cal N}_0 \delta {\cal A}_i \dot{\delta \cal B}_i
-\frac{1}{a^2}W_i \delta {\cal A}_i \right]\,.
\label{LVM}
\ee
Since the quantities $W_i, \delta {\cal A}_i, \delta {\cal B}_i$ appear 
only in the matter action, we can vary the action (\ref{LVM}) with 
respect to these perturbations independently of the full 
second-order action. 
Variation of Eq.~(\ref{LVM}) in terms of $W_i$ leads to 
\be
W_i=\frac{{\cal N}_0 (\delta {\cal A}_i-\rho_{M,n}V_i)}{\rho_{M,n}}\,.
\label{Wiex}
\ee
Substituting Eq.~(\ref{Wiex}) into Eq.~(\ref{LVM}) and varying the action 
with respect to $\delta {\cal A}_i$, we obtain $\delta {\cal A}_i$ 
in the form (\ref{delAi}) with 
\be
u_i=V_i-a^2\dot{\delta {\cal B}}_i\,.
\ee
Varying the action with respect to $\delta {\cal B}_i$, we find 
\be
\rho_{M,n}\,u_{i}
=\frac{\rho_M+P_M}{n_0}\,u_{i}
=\mathrm{constant}\,,
\label{conser}
\ee
which is associated with the conservation of the energy-momentum tensor 
of the perfect fluid. This is the same relation as that in GR and it states that
the perturbation $u_i$ is non-dynamical. 
Finally, the second-order matter action reduces to 
\be
(S_V^{(2)})_M=\sum_{i=1}^{2} \int dtd^3x\,
\frac{a}{2} \left[ \left( \rho_M+P_M \right)
\left( V_i-a^2\dot{\delta {\cal B}}_i \right)^2
-\rho_M V_i^2 \right]\,,
\label{LVM2}
\ee
with the conservation relation (\ref{conser}) for the four velocity
$u_i=V_i-a^2\dot{\delta {\cal B}}_i$.

We sum up the second-order action originating from $\int d^4 x \sqrt{-g}{\cal L}$ 
with Eq.~(\ref{LVM2}). In doing so, it is convenient to  introduce 
the following combination
\be
Z_i= E_i+\phi(t)\,V_i\,.
\ee
Let us consider the perturbations in Fourier space with $k=|{\bm k}|$.
Variation of the total second-order action $S_V^{(2)}$ with 
respect to $V_{i}$ leads to
\be
\frac{q_T}{2}
\frac{k^{2}}{a^{2}}V_{i}=-(\rho_M+P_M)u_{i}-\phi\,
\left(G_{4,X}-\frac12 G_{5,X}H \phi \right)\,\frac{k^{2}}{a^{2}}Z_{i}\,.
\label{vecre}
\ee
The metric perturbation $V_{i}$ follows
a different dynamics from that in GR (where $G_{4,X}=0=G_{5,X}$). 
Integrating out the non-dynamical fields $u_i$ and $V_i$, 
we obtain the action for the dynamical field $Z_i$.
It follows that there are two dynamical degrees of freedom
$Z_1$ and $Z_2$ for the vector mode.

Taking the small-scale limit ($k \to \infty$) in Eq.~(\ref{vecre}) and using the
background equations of motion, the resulting second-order 
action for the perturbations $Z_i$ in Fourier space reads
\be
S_V^{(2)} \simeq \sum_{i=1}^{2} \int dt\,d^3 x\,
\frac{a q_V}{2} \left( \dot{Z}_i^2+\frac{k^2}{a^2}c_V^2
Z_i^2 \right)\,,
\label{SVac}
\ee
where
\be
q_V= 1-2c_{2}G_{4,X}-2d_2H\phi\,G_{5,X}\,,
\label{qv}
\ee
and the vector propagation speed squared can be written as 
\be
c_{V}^{2}=1+\frac{\phi^2(2G_{4,X}-G_{5,X}H \phi)^2}
{2q_Tq_V}+\frac{d_2G_{5,X}(H\phi-\dot{\phi})}{q_V}\,.
\label{cv}
\ee
In Eq.~(\ref{SVac}) we have ignored the contribution of 
an effective mass term $m_V^2Z_i^2$ relative to the second term, 
which can be justified in the limit $k \to \infty$ with $c_V \neq 0$.

The sign of the coefficient $q_V$ characterizes the no-ghost
condition, such that the vector ghost is absent for $q_V>0$.
If $c_2=d_2=0$, then this condition is
automatically satisfied.
We recall that the terms containing $c_2$ and $d_2$ are 
associated with the pure vector contributions 
in the original action (\ref{Lag}), 
which affect the no-ghost condition
for the vector mode.

If the vector propagation speed squared $c_V^2$ is
positive, the small-scale Laplacian instability can be avoided.
For the theories with $G_5=0$ or the theories with
$d_2=0$ and $G_5 \neq 0$, we have that $c_V^2>1$
under the no-ghost conditions of tensor and vector
perturbations ($q_T>0$ and $q_V>0$).

\subsection{Scalar perturbations}

For scalar perturbations, we consider the single perfect fluid with
the background energy density $\rho_M$ and
the pressure $P_M$ given by Eq.~(\ref{PM}).
We make the field redefinitions
\ba
\delta J &=&
\frac{a^{3}}{\rho_{M,n}}\,\delta \rho_{M}
=\frac{n_0\,a^3}{\rho_{M}+P_{M}}\,\delta \rho_{M}\,,\\
\chi_{V}&=& \psi-\phi(t) \chi\,,
\ea
where $\delta\rho_{M}$ is the matter density perturbation.
First, we can integrate out the Lagrange multiplier $\delta j$
by means of its own equation of motion:
\be
\frac{\partial \delta j}{{\mathcal N}_0}
=-\partial v-\partial \chi\,.
\ee

Then, the second-order action for scalar perturbations is given by 
$S_{S}^{(2)}=\int dt d^3x\,L_{S}^{(2)}$, where
\ba
L_{S}^{(2)} & = & a^{3}\,\Biggl\{-\frac{n_0 \rho_{M,n}}{2}\,\frac{(\partial v)^{2}}{a^{2}}
+\left[ n_0\rho_{M,n}\,\frac{\partial^{2}\chi}{a^2}-\dot{\delta\rho}_M
-3H\left(1+\frac{n_0\rho_{M,nn}}{\rho_{M,n}} \right)\,\delta\rho_M \right]v
-\frac{\rho_{M,nn}}{2\rho_{M,n}^{2}}(\delta \rho_M)^{2}-\alpha\,\delta\rho_M \nonumber \\
 &  & {}-w_{3}\,\frac{(\partial\alpha)^{2}}{a^{2}}+w_{4}\alpha^{2}
 -\left[(3Hw_{1}-2w_{4})\frac{\delta\phi}{\phi}-w_{3}\,\frac{\partial^{2}(\delta\phi)}
 {a^{2}\phi}-w_{3}\,\frac{\partial^{2}\dot{\psi}}{a^{2}\phi}
 +w_{6}\,\frac{\partial^{2}\psi}{a^{2}}\right] \alpha\nonumber \\
 &  & {}-\frac{w_{3}}{4}\,\frac{(\partial\delta\phi)^{2}}{a^{2}\phi^{2}}
 +w_{5}\,\frac{(\delta\phi)^{2}}{\phi^{2}}
 -\left[\frac{(w_{6}\phi+w_{2})\psi}{2}-\frac{w_{3}}{2}\dot{\psi}\right]
 \frac{\partial^{2}(\delta\phi)}{a^{2}\phi^{2}}\nonumber \\
 &  & {}-\frac{w_{3}}{4\phi^{2}}\,\frac{(\partial\dot{\psi})^{2}}{a^{2}}+\frac{w_{7}}{2}\,
 \frac{(\partial\psi)^{2}}{a^{2}}
 +\left(w_{1}\alpha+\frac{w_{2}\delta\phi}{\phi}\right)\frac{\partial^{2}\chi}{a^{2}}\Biggr\}\,,
\label{sscalar}
\ea
with the short-cut notations
\ba
w_{1} & = & {H}^{2} {\phi}^{3} (G_{{5,X}}+{\phi}^{2}G_{{5,{\it XX}}})
-4\,H(G_{{4}}+{\phi}^{4}G_{{4,{\it XX}}})-{\phi}^{3}G_{{3,X}}\,,\nonumber \\
w_{2} & = & w_1+2Hq_T\,,\nonumber \\
w_{3} & = & -2\,{\phi}^{2}q_V\,,\nonumber \\
w_{4} & = & \frac{1}{2}{H}^{3}\phi^{3}(9G_{{5,X}}-\phi^{4}G_{{5,{\it XXX}}})
-3\,H^{2} (2G_{{4}}+2\phi^{2}G_{{4,X}}+\phi^{4}G_{{4,{\it XX}}}-\phi^{6}G_{{4,{\it XXX}}}) \nonumber\\
 &  & {}-\frac{3}{2}\,H\phi^{3}(G_{{3,X}}-\phi^{2}G_{{3,{\it XX}}})
 +\frac{1}{2}\,\phi^{4}G_{{2,{\it XX}}}\,,\nonumber\\
w_{5} & = & w_{4}-\frac{3}{2}\,H(w_{1}+w_{2})\,, \nonumber\\
w_{6} & = & -\phi\,\left[{H}^{2}\phi(G_{{5,X}}-{\phi}^{2}G_{{5,{\it XX}}})
-4\,H(G_{{4,X}}-{\phi}^{2}G_{{4,{\it XX}}})+\phi G_{{3,X}}\right]\,, \nonumber\\
w_{7} & = & 2(H\phi G_{{5,X}}-2G_{{4,X}}) \dot{H}
+\left[H^{2}(G_{{5,X}}+{\phi}^{2}G_{{5,{\it XX}}})-4\,H\phi\,G_{{4,{\it XX}}}-G_{{3,X}}\right] \dot{\phi}\,.
\ea
The Lagrangian (\ref{sscalar}) does not contain the time derivatives
of $\alpha, \chi, \delta \phi, v$, so these four fields are non-dynamical.
Varying the action $S_S^{(2)}$ with respect to $\alpha, \chi, \delta \phi, v$,
we obtain the following equations of motion in Fourier space respectively:
\ba
& &
\delta \rho_M-2w_4 \alpha+\left( 3Hw_1-2w_4 \right)\frac{\delta \phi}{\phi}
+\frac{k^2}{a^2} \left( w_3\frac{\delta \phi}{\phi}+w_3 \frac{\dot{\psi}}{\phi}
-w_6 \psi+w_1 \chi+2w_3 \alpha \right)=0\,,\\
& &
n_0\rho_{M,n} v+
w_1 \alpha+\frac{w_2}{\phi} \delta \phi=0\,,\\
& &
\left( 3Hw_1-2w_4 \right)\alpha-2w_5 \frac{\delta \phi}{\phi}
+\frac{k^2}{a^2} \left( w_3\alpha+\frac{w_3 \delta \phi}
{2\phi}-\frac{w_6\phi+w_2}{2\phi}\psi+\frac{w_3\dot{\psi}}{2\phi}+w_2 \chi \right)=0\,,\\
& &
\dot{\delta \rho}_M+3H\left(1+\frac{n_0\rho_{M,nn}}{\rho_{M,n}} \right) \delta \rho_M
+\frac{k^2}{a^2}n_0\rho_{M,n} \left( \chi+v \right)=0\,.
\ea

On using these equations, we can eliminate the non-dynamical fields from
the second-order Lagrangian $L_S^{(2)}$.
After integrations by parts, we finally obtain a reduced Lagrangian
for two dynamical fields $\psi$ and $\delta\rho_M$ in the form
\be
L_{S}^{(2)}=a^{3}\left( \dot{\vec{\mathcal{X}}}^{t}{\bm K}
\dot{\vec{\mathcal{X}}}
+\frac{k^2}{a^2}\vec{\mathcal{X}}^{t}{\bm G}
\vec{\mathcal{X}}
-\vec{\mathcal{X}}^{t}{\bm M}
\vec{\mathcal{X}}
-\vec{\mathcal{X}}^{t}{\bm B}
\dot{\vec{\mathcal{X}}}
\right) \,,
\label{L2mat}
\ee
where ${\bm K}$, ${\bm G}$, ${\bm M}$, ${\bm B}$
are $2 \times 2$ matrices, and the vector field
$\vec{\mathcal{X}}$ is defined by
\be
\vec{\mathcal{X}}^{t}=\left( \psi, \delta\rho_M/k \right) \,.
\ee
The matrix ${\bm M}$ is related with the masses of two
scalar modes, which do not contain the $k^2/a^2$ term.
In the following we shall take the small-scale limit
($k \to \infty$), in which the second term on the r.h.s. 
of Eq.~(\ref{L2mat}) dominates over the third 
and fourth terms.

The matrix ${\bm K}$ is associated with the kinetic terms
of two scalar modes.
Provided that the two eigenvalues of ${\bm K}$ are positive,
the scalar ghosts are absent.
The no-ghost condition for the matter perturbation $\delta \rho_M$ 
is trivially satisfied for $\rho_M+P_M>0$.
For the perturbation $\psi$, the quantity related with
the no-ghost condition corresponds
to\footnote{The fluid description (\ref{Spf}) based on the matter perturbation 
$\delta \rho_M$ is convenient in that the off-diagonal components of the matrix ${\bm K}$
vanish. This is not generally the case for the k-essence description of the perfect
fluid, see e.g., Refs.~\cite{kesoff}.}
\be
Q_{S}=\frac{a^{3}H^2q_Tq_S}{\phi^{2}(w_{1}-2w_{2})^{2}}\,,
\label{Qsge}
\ee
where
\be
q_S=3w_{{1}}^{2}+4q_Tw_4\,.
\label{qs}
\ee
Under the condition $q_T>0$, the scalar ghost does not
arise for $q_S>0$.

For large $k$, the Lagrangian (\ref{L2mat}) leads
to the dispersion relation
\be
{\rm det} \left( \omega^2 {\bm K}-\frac{k^2}{a^2}
{\bm G} \right)=0\,,
\label{deteq}
\ee
where $\omega$ is a frequency.
Defining the scalar propagation speed $c_s$ according to
the relation $\omega^2=c_s^2\,k^2/a^2$, there are two solutions
to Eq.~(\ref{deteq}).
One of them is the matter propagation speed squared given by
\begin{equation}
c_M^2=
\frac{n_0\rho_{M,nn}}{\rho_{M,n}}\,,
\end{equation}
which is the same value as that of GR.
Another is the propagation speed squared of the perturbation $\psi$,
which is given by
\ba
c_{S}^{2} &=&
 \frac{1}{\Delta}\,\Big\{2\,w_{{2}}^{2}w_{{3}}(\rho_{M}+P_{M})
-w_3(w_1-2w_2)\left[w_1w_2+\phi(w_1-2w_2)w_6 \right]
\left(\dot{\phi}/\phi-H\right)
- w_3\left( 2w_2^2\dot{w}_1-w_1^2\dot{w}_2 \right)\nonumber\\
 &  & {}+\phi \left(w_1-2w_2 \right)^2 \left[w_3\dot{w}_6
 +\phi(2w_3w_7+w_6^2) \right]
 +w_1w_2\left[ w_1w_2+(w_1-2w_2)(2\phi\,w_6-w_3\dot{\phi}/\phi)
 \right]\Big\}\,,\label{cs}
\ea
where
\be
\Delta \equiv 8H^2 \phi^2 q_Tq_Vq_S\,.
\ee
If the numerator of Eq.~(\ref{cs}) is positive, then the Laplacian
instability of scalar perturbations is absent under the three
no-ghost conditions $q_T>0,q_V>0$, and $q_S>0$.

While we have focused on the single-fluid case, the results given in
Eqs.~(\ref{Qsge}) and (\ref{cs}) are valid for the multi-fluid case
by dealing with $\rho_M$ and $P_M$ as the energy density
and the pressure of the total fluid, respectively.
This situation is analogous to what happens in scalar
Horndeski theories \cite{extended1}.
In summary, besides the matter fluid, the number of propagating
degrees of freedom is five (two tensors, two vectors, one scalar),
where three of them (two vectors and one scalar) originate from 
the massive vector field. 
The six quantities $q_T,q_V,q_S$ and $c_T^2,c_V^2,c_S^2$ need to
be positive to avoid the appearance of ghosts and instabilities
for tensor, vector, scalar perturbations.

\section{General discussions for de Sitter and
early-Universe stabilities}
\label{genedissec}

We discuss no-ghost and stability conditions derived in Sec.~\ref{spsec}
in the two stages:
(i) de Sitter fixed point, and (ii) early cosmological epochs
(radiation/matter eras).
The functional forms of $G_{2,3,4,5}$
are not specified in this section.

\subsection{de Sitter fixed point}

The de Sitter solutions, which are characterized by
$\dot{\phi}=0$ and $\dot{H}=0$, can exist for
the branch satisfying Eq.~(\ref{be3d}).
We can solve Eqs.~(\ref{be1}) and (\ref{be3d}) to express
$G_{4,X}$ and $G_{4,XX}$ in terms of other quantities
like $G_2$.
Since $\rho_M=0$ at the de Sitter solution, the quantities
$q_T$ and $c_T^2$ reduce, respectively, to
\be
q_T=-\frac{G_2-H^3\phi^3G_{5,X}}{3H^2}\,,
\qquad
c_T^2 = \frac{2G_4}{q_T}\,.
\label{ctds}
\ee
Provided that $G_4>0$, the tensor ghost and instability
are absent for $q_T>0$, i.e.,
\be
G_2-H^3\phi^3G_{5,X}<0\,.
\label{qT2}
\ee

In the absence of the Lagrangian ${\cal L}_5$ (i.e., $G_5=0$),
the condition (\ref{qT2}) simply reduces to $G_2<0$.
This sign is opposite to that of standard Proca theories
characterized by the function $G_2=m^2X$.
Naively one may think that this gives rise to tachyons, but
we need to caution that the effective mass $m_V$ of the vector
field in the presence of gravity is generally different from its bare mass.

The stability condition for super-horizon modes could be a subtle
issue since the expression for the squared mass depends on e.g., normalization
of variables. So far, we have defined the variables $E_i$ and $V_i$ in
the comoving coordinate basis $dx^i$. Alternatively, one could define the
variables $\tilde{E}_i$ and $\tilde{V}_i$ in the
background tetrad basis $a(t)dx^i$ since it is the tetrad components rather than
the comoving coordinate components that are relevant for local nonlinear interactions.
The new set of variables is related to the original set of variables as
$E_i=a\tilde{E}_i$ and $V_i=a\tilde{V}_i$, and thus the new variables
may stay constant or decay even if the original variables grow exponentially.
In terms of the new variable $\tilde{E}_i$, the quadratic action for the
vector perturbations around de Sitter backgrounds is simply given by
\be
S_V^{(2)}=\sum_{i=1}^{2} \int dt\,d^3 x\,
\frac{a^3\, q_V}{2} \left[ \dot{\tilde{E}}_i^2-\frac{c_V^2}{a^2}
(\partial \tilde{E}_i)^2 -2H^2 \tilde{E}_i^2 \right]\,.
\ee
This expression holds for all $k$ (i.e., without the need for taking the limit $k\to\infty$).
Since the effective mass squared $m_V^2=2H^2$ of the vector mode
is positive, this implies the absence of tachyonic instability on the de Sitter
solution. For $m$ of the order of the today's Hubble parameter $H_0$,
the vector mass term does not affect the dynamics of
perturbations deep inside the Hubble radius ($k/a \gg H$)
even in the early cosmological epoch.

If the Lagrangian ${\cal L}_5$ is present, it is possible to
satisfy the condition (\ref{qT2}) even for $G_2>0$.
In this case, however, both $Q_T$ and $c_T^2$ are
negative for $G_5 \to 0$, so the ghost and the instability 
arise in this limit. We shall focus on the theories with
$G_2<0$ to ensure the stability of tensor perturbations
in the limit $G_5 \to 0$.

{}From Eqs.~(\ref{qv}) and (\ref{cv}) the quantities
associated with vector perturbations
at the de Sitter point are given by
\ba
q_V &=&
1-\frac{c_2(G_2+6H^2 G_4
+2H^3\phi^3G_{5,X})
+6d_2H^3\phi^3G_{5,X}}{3H^2\phi^2}\,,\\
c_V^2 &=&
1+\frac{(G_2+6H^2G_4-H^3\phi^3G_{5,X})^2}
{18H^4\phi^2 q_T q_V}
+\frac{d_2H\phi G_{5,X}}{q_V}\,.
\ea
For the theories and backgrounds
with $d_2 H\phi G_{5,X}>0$, it follows
that $c_V^2>1$ under the two no-ghost conditions
$q_T>0$ and $q_V>0$.
For $d_2 H\phi G_{5,X}<0$,
the condition for avoiding the Laplacian
instability corresponds to
\be
18H^4\phi^2 q_T
\left( q_V+d_2H\phi G_{5,X}\right)
+(G_2+6H^2G_4-H^3 \phi^3 G_{5,X})^2>0\,.
\ee

For scalar perturbations, all the terms with the time-derivatives
in Eq.~(\ref{cs}) vanish at the de Sitter solution. Then $c_S^2$
can be simply written as
$c_S^2=[(w_1-2w_2)\phi w_6+w_1w_2]
[(w_1-2w_2)(\phi w_6+H w_3)+w_1w_2]/\Delta$.
By using the functions $G_{2,3,4,5}$ and their derivatives,
Eqs.~(\ref{qs}) and (\ref{cs}) reduce, respectively, to
\ba
q_S &=&
\frac{1}{3H^2}\Big[
\left\{2G_2+2\phi^2G_{2,X}+3 H\phi^3G_{3,X}
+H^3\phi^3(G_{5,X}+\phi^2G_{5,XX})\right\}^2
+2(G_2-H^3\phi^3G_{5,X})\big\{G_2-\phi^2G_{2,X}
\notag\\
&&
-\phi^4G_{2,XX}-3H\phi^5G_{3,XX}
+6H^2(3G_4-\phi^6G_{4,XXX})
-H^3\phi^3(4G_{5,X}-\phi^2G_{5,XX}-\phi^4G_{5,XXX})\big\}\Big]\,,
\label{qsds}\\
c_S^2 &=& \frac{\xi_{S1}(\xi_{S1}-\xi_{S2})}{648 H^6 \phi^2 q_Tq_Vq_S}\,,
\label{csds}
\ea
where
\ba
\xi_{S1} &\equiv&
\left[2G_2-2\phi^2G_{2,X}-3H\phi^3G_{3,X}
-H^3\phi^3(5G_{5,X}+\phi^2G_{5,XX})\right]
\left[4G_2+2\phi^2G_{2,X}+3H\phi^3G_{3,X}
+24H^2G_4\right.\notag\\
&&\left.-H^3\phi^3(G_{5,X}-\phi^2G_{5,XX})\right]
+\phi^2\left[2G_{2,X}+3H\phi G_{3,X}
+H^3\phi(3G_{5,X}+\phi^2G_{5,XX})\right]
\left[2G_2+2\phi^2G_{2,X}\right.\notag\\
&&\left.+3H\phi^3G_{3,X}
+H^3\phi^3(G_{5,X}+\phi^2G_{5,XX})\right]
\,,\label{xi1}\\
\xi_{S2} &\equiv&
6H^2\phi^2\left[2G_2-2\phi^2G_{2,X}-3H\phi^3G_{3,X}
-H^3\phi^3(5G_{5,X}+\phi^2G_{5,XX})\right]\notag\\
&&
\times\left[1-\frac{c_2(G_2+6H^2 G_4
+2H^3\phi^3G_{5,X})
+6d_2H^3\phi^3G_{5,X}}{3H^2\phi^2}\right]
\,.\label{xi2}
\ea
Under the two conditions $q_T>0$ and $q_V>0$,
the scalar ghost and the instability are absent for
\be
q_S>0\,,\qquad \xi_{S1}(\xi_{S1}-\xi_{S2})>0\,.
\ee
Unless we specify some functional forms of $G_{2,3,4,5}$,
it is difficult to derive concrete constraints on these functions
from the general expressions (\ref{qsds})-(\ref{csds}).

\subsection{Early cosmological epochs}

Let us proceed to the discussion of no-ghost and stability
conditions during radiation and matter eras.
In doing so, we assume that the early cosmological evolution
is close to that of GR with the function
$G_4=M_{\rm pl}^2/2+g_4$ satisfying the condition
$g_4 \ll M_{\rm pl}^2/2$.
The background Eqs.~(\ref{be1}) and (\ref{be2})
can be expressed in the following forms
\ba
& &
3M_{\rm pl}^2 H^2=\rho_M+\rho_{\rm DE}\,,
\label{Fri1}\\
& &
M_{\rm pl}^2 ( 3H^2+2\dot{H})
=-P_M-P_{\rm DE}\,,
\label{Fri2}
\ea
where
\ba
\rho_{\rm DE} &=&
-G_2+G_{2,X}\phi^2+3G_{3,X}\phi^3 H
-6g_4H^2+6\phi^2H^2 (2G_{4,X}+G_{4,XX}\phi^2)
-H^3 G_{5,XX}\phi^5-5H^3G_{5,X} \phi^3,
\label{rhode}\\
P_{\rm DE} &=&
G_2-\dot{\phi}\phi^2G_{3,X}+2g_4\,(3H^2+2\dot{H})
-2\phi G_{4,X}\, ( 3\phi H^2 +2\dot{\phi}H
+2\phi \dot{H} )-4H\dot{\phi}\phi^3G_{4,XX}\nonumber\\
&& {}+\dot\phi\phi^4H^2G_{5,XX}+G_{5,X}\phi^2H(2\phi\dot H+2\phi H^2+3\dot\phi H)\,.
\label{Pde}
\ea

In the early cosmological epoch, the energy density
$\rho_{\rm DE}$ and the pressure $P_{\rm DE}$
of the ``dark'' component are sub-dominant to
$\rho_M$ and $P_M$, respectively, so that
$|\rho_{\rm DE}| \ll M_{\rm pl}^2 H^2$ and
$|P_{\rm DE}| \ll M_{\rm pl}^2 H^2$.
It is then natural to consider a situation in which each term 
in Eqs.~(\ref{rhode}) and (\ref{Pde}) is much smaller 
than the order of $M_{\rm pl}^2 H^2$.
In this case, the terms $\phi^2 G_{4,X}$,
$H \phi^3 G_{5,X}$, and $\phi^2 \dot{\phi}G_{5,X}$
in $q_T$ and $c_T^2$ are much smaller than
the order of $2G_{4} \simeq M_{\rm pl}^2$, so the quantities
(\ref{qT}) and (\ref{cT}) are approximately given by
\be
q_T \simeq M_{\rm pl}^2\,,\qquad
c_T^2 \simeq 1\,.
\ee
Hence there are neither ghosts nor instabilities
for tensor perturbations.

For vector perturbations, even if each term in Eqs.~(\ref{rhode}) 
and (\ref{Pde}) is much smaller than the order of $M_{\rm pl}^2 H^2$, 
this does not necessarily imply that the terms $c_2G_{4,X}$ and 
$d_2H\phi G_{5,X}$ in $q_V$ are much less than 1.
If the conditions
\be
|c_2 G_{4,X}| \ll 1\,,\qquad
|d_2 H\phi G_{5,X}| \ll 1\,,\qquad
|\phi| \lesssim M_{\rm pl}
\label{qvearly}
\ee
hold in the early cosmological epoch with
$|c_2|, |d_2|$ of the order of unity,
$q_V$ and $c_V^2$ approximately reduce to
\be
q_V \simeq 1\,,\qquad
c_V^2 \simeq 1\,.
\ee
In this case, the vector ghost and stability conditions
are automatically satisfied.

It should be noted that, even if the conditions (\ref{qvearly}) are 
violated in the early cosmological epoch, one can avoid the vector
ghost for  $c_2 G_{4,X}<0$ and $d_2 H\phi G_{5,X}<0$.
In addition, $c_V^2$ is larger than 1
for $d_2G_{5,X}(H\phi-\dot{\phi})>0$.
Hence, the three conditions in Eq.~(\ref{qvearly}) 
are sufficient but not necessary, so
there are other cases in which both $q_V$ and $c_V^2$
are positive.

The quantities $q_S$ and $c_S^2$ of scalar perturbations
contain the third-order derivatives $G_{4,XXX}$
and $G_{5,XXX}$ with respect to $X$.
These terms do not appear in the background
equations of motion. Hence we need to impose some conditions
on such derivatives to derive analytic expression
of $Q_S$ and $c_S^2$. Moreover, unless the functional
forms of $G_{2,3,4,5}$ are specified, one cannot extract
detailed information for each function due to the complexities
of no-ghost and stability conditions of scalar
perturbations. In Sec.~\ref{covasec} we shall compute
the quantities $Q_S$ and $c_S^2$ for concrete 
models from the radiation era to the late-time de Sitter solution.

\section{Application to concrete models} 
\label{covasec}

In this section, we search for viable dark energy models
in the framework of generalized Proca theories given by
the action (\ref{Lag}).
{}From Eq.~(\ref{be3d}) the field $\phi$ depends on
the Hubble parameter $H$.
To realize the situation in which the energy density of
$\phi$ starts to dominate over the background matter
densities at the late cosmological epoch, the amplitude
of the field $\phi$ should grow with the decrease of $H$.
For this purpose, we assume that $\phi$ is related with
$H$ according to the relation
\be
\phi^p \propto H^{-1}\,,
\label{phiH}
\ee
where $p$ is a positive constant.
We shall later justify this ansatz by showing that all background 
equations of motion are satisfied and that all stability conditions 
for linear perturbations are fulfilled.
Hereafter, without loss of generality, we shall focus on 
the branch in which $\phi$ is positive.

For realizing the solution (\ref{phiH}), the functions
$G_{2,3,4,5}$ in Eq.~(\ref{be3d}) should contain
the power-law dependence of $X$ in the forms
\be
G_2(X)=b_2 X^{p_2}\,,\qquad
G_3(X)=b_3X^{p_3}\,,\qquad
G_4(X)=\frac{M_{\rm pl}^2}{2}+b_4X^{p_4}\,,\qquad
G_5(X)=b_5X^{p_5}\,,
\label{G2345}
\ee
where $M_{\rm pl}$ is the reduced Planck mass, 
$b_{2,3,4,5}$ are constants, and the powers $p_{3,4,5}$ 
are given by
\be
p_3=\frac12 \left( p+2p_2-1 \right)\,,\qquad
p_4=p+p_2\,,\qquad
p_5=\frac12 \left( 3p+2p_2-1 \right)\,.
\label{p345}
\ee
The vector Galileon \cite{Heisenberg,Tasinato} corresponds to
the powers $p_2=1$ and $p=1$, so the field $\phi$ has the dependence $\phi \propto H^{-1}$.
Thus the analysis based on the assumption (\ref{phiH})
is general in that it encompasses the vector Galileon as a specific case.

\subsection{Dynamical analysis of the background}\label{backgrounddynamics}

Let us first consider the background dynamics for the theories
given by the functions (\ref{G2345}) with
the powers (\ref{p345}). For the matter sector
we take into account non-relativistic matter (energy density
$\rho_m$ and pressure $P_m=0$) and radiation
(energy density $\rho_r$ and pressure $P_r=\rho_r/3$), which
obey the continuity equations
$\dot{\rho}_m+3H\rho_m=0$ and
$\dot{\rho}_r+4H\rho_r=0$ respectively.
In this case, we have $\rho_M=\rho_m+\rho_r$ and
$P_M=\rho_r/3$ in Eqs.~(\ref{Fri1}) and (\ref{Fri2}).

To study the dynamical system of the background,
it is convenient to introduce the following
density parameters
\be
\Omega_{r} \equiv
\frac{\rho_r}{3M_{\rm pl}^2H^2}\,,
\qquad
\Omega_{m} \equiv
\frac{\rho_m}{3M_{\rm pl}^2H^2}\,,
\qquad
\Omega_{\rm DE} \equiv
1-\Omega_{r}-\Omega_{m}\,,
\ee
and the following dimensionless quantities
\be
y \equiv \frac{b_2\phi^{2p_2}}
{3M_{\rm pl}^2 H^2\,2^{p_2}}\,,\qquad
\beta_i \equiv
\frac{p_ib_i}{2^{p_i-p_2}p_2b_2}
\left( \phi^p H \right)^{i-2}\,,
\label{ydef}
\ee
where $i=3,4,5$. Due to the relation (\ref{phiH}),
$\beta_i$'s are constants.
{}From Eq.~(\ref{be3d}) it follows that
\be
1+3\beta_3+6(2p+2p_2-1)\beta_4
-(3p+2p_2)\beta_5=0\,.
\label{mueq}
\ee
In the following, we employ this relation to express $\beta_3$
in terms of $\beta_4$ and $\beta_5$.

On using Eq.~(\ref{be1}), we find that the dark energy
density parameter is related with the quantity $y$, as
\be
\Omega_{\rm DE}=
\frac{\beta y}{p_2(p+p_2)}\,,
\label{OmegaDE}
\ee
where
\be
\beta \equiv -p_2(p+p_2)(1+4p_2 \beta_5)
+6p_2^2(2p+2p_2-1)\beta_4\,.
\label{betadef}
\ee
Differentiating Eq.~(\ref{be3d}) with respect to $t$
and using Eq.~(\ref{be2}), we can solve for
$\dot{\phi}$ and $\dot{H}$.
Taking the derivatives of
$\Omega_{\rm DE}$ and $\Omega_{r}$ with respect to
${\cal N} \equiv \ln a$ (denoted as a prime),
we obtain the following dynamical equations of motion
\ba
\Omega_{\rm DE}' &=& \frac{(1+s)\Omega_{\rm DE}
(3+\Omega_r-3\Omega_{\rm DE})}
{1+s\,\Omega_{\rm DE}}\,,
\label{dx1} \\
\Omega_{r}' &=& -\frac{\Omega_r [ 1-\Omega_r
+(3+4s)\Omega_{\rm DE}]}
{1+s\,\Omega_{\rm DE}}\,,
\label{dx2}
\ea
where
\be
s \equiv \frac{p_2}{p}\,.
\ee
The matter density parameter $\Omega_m=1-\Omega_{\rm DE}-\Omega_{r}$
is known by solving Eqs.~(\ref{dx1}) and (\ref{dx2}) for
$\Omega_{\rm DE}$ and $\Omega_r$, respectively, with their given
initial conditions. We shall prevent $\Omega_{\rm DE}$ 
(or $\Omega_{r}$) from diverging in the interval
$0\leq\Omega_{\rm DE}\leq 1$, by imposing $1+s>0$.

The effective equation of state of the system, which is
defined by $w_{\rm eff} \equiv -1-2\dot{H}/(3H^2)$, reads
\be
w_{\rm eff}=
\frac{\Omega_r-3(1+s)\Omega_{\rm DE}}
{3(1+s\,\Omega_{\rm DE})}\,.
\label{weff}
\ee
We define the dark energy equation of state as
$w_{\rm DE} \equiv P_{\rm DE}/\rho_{\rm DE}$,
where $\rho_{\rm DE}$ and $P_{\rm DE}$ are
given by Eqs.~(\ref{rhode}) and (\ref{Pde})
respectively. Then, it follows that
\be
w_{\rm DE}=
-\frac{3(1+s)+s\,\Omega_r}
{3(1+s\,\Omega_{\rm DE})}\,.
\label{wde}
\ee

For the dynamical system given by Eqs.~(\ref{dx1})
and (\ref{dx2}), there are the following three fixed points:
\ba
& &{\rm (a)}~~(\Omega_{\rm DE},\Omega_{r})=(0,1) \qquad \quad ({\rm radiation~point})\,,\\
& &{\rm (b)}~~(\Omega_{\rm DE},\Omega_{r})=(0,0) \qquad \quad ({\rm matter~point})\,,\\
& &{\rm (c)}~~(\Omega_{\rm DE},\Omega_{r})
=(1,0) \qquad \quad ({\rm de~Sitter~point})\,.
\ea
At each fixed point, the equations of state (\ref{weff}) and
(\ref{wde}) reduce to
\ba
& &{\rm (a)}~~
w_{\rm eff}=\frac13\,,\qquad w_{\rm DE}
=-1-\frac{4}{3}s\,,\label{casea}\\
& &{\rm (b)}~~
w_{\rm eff}=0\,,\qquad w_{\rm DE}=-1-s\,,\label{caseb}\\
& &{\rm (c)}~~
w_{\rm eff}=-1\,,\qquad w_{\rm DE}=-1\,.\label{casec}
\ea

{}From Eqs.~(\ref{dx1}) and (\ref{dx2}) we obtain the relation
\be
\frac{\Omega_{\rm DE}'}{\Omega_{\rm DE}}
=(1+s) \left( \frac{\Omega_{r}'}{\Omega_{r}}+4 \right)\,.
\ee
Integrating this equation gives
\be
\frac{\Omega_{\rm DE}}{\Omega_{r}^{1+s}}
=C a^{4(1+s)}\,,
\label{Omere}
\ee
where $C$ is an integration constant.
The constant $C$ can be fixed as
$Ca_0^{4(1+s)}=(\Omega_{\rm DE}/\Omega_{r}^{1+s})_0$, 
where the lower subscript ``0'' represents the today's value.
Since the ratio between $\Omega_m$
and $\Omega_r$ is given by
$\Omega_m/\Omega_r=(a/a_0)\,(\Omega_m/\Omega_r)_0$, 
elimination of the scale factor $a$ from Eq.~(\ref{Omere})
leads to
\be
\frac{\Omega_{\rm DE}\,\Omega_r^{3(1+s)}}
{(1-\Omega_{\rm DE}-\Omega_r)^{4(1+s)}}
=\left( \frac{\Omega_{\rm DE}}{\Omega_r^{1+s}}
\right)_0
\left( \frac{\Omega_{r}}{1-\Omega_{\rm DE}-\Omega_r}
\right)_0^{4(1+s)}\,,
\ee
which corresponds to the trajectory of solutions
in the $(\Omega_{\rm DE}, \Omega_{r})$ plane.
For given values of $\Omega_{{\rm DE}0}$ and
$\Omega_{r0}$, the cosmological trajectory
is fixed.

In the radiation-dominated epoch ($\Omega_r \simeq 1$),
the relation (\ref{Omere}) shows that, for $s>-1$,
the dark energy density parameter grows as
$\Omega_{\rm DE} \propto a^{4(1+s)}$.
In fact, this is consistent with the limits $\Omega_r \to 1$
and $\Omega_{\rm DE} \ll 1$ in Eq.~(\ref{dx1}), i.e.,
$\Omega_{\rm DE}' \simeq 4(1+s) \Omega_{\rm DE}$.
Since $\Omega_{\rm DE} \propto y \propto
\phi^{2p_2}/H^2$, the evolution of $\phi$
during the radiation era ($H \propto t^{-1}$ and
$a \propto t^{1/2}$) is given by $\phi \propto a^{2s/p_2}$.
In summary, around the radiation point (a),
the dark energy density parameter and the field
$\phi$ evolve as
\be
\Omega_{\rm DE} \propto t^{2(p+p_2)/p}\,,
\qquad \phi \propto t^{1/p}\,.
\label{Omedephi}
\ee

In the matter-dominated epoch ($\Omega_m \simeq 1$)
the radiation density parameter $\Omega_r$ decreases as
$\Omega_r \propto a^{-1}$, so Eq.~(\ref{Omere})
implies that the dark energy density parameter
evolves as $\Omega_{\rm DE} \propto a^{3(1+s)}$.
This relation can be also confirmed by taking the
limits $\Omega_r \ll 1$ and $\Omega_{\rm DE} \ll 1$
in Eq.~(\ref{dx1}). On using the properties
$H \propto t^{-1}$ and $a \propto t^{2/3}$,
it follows that the time dependence of
$\Omega_{\rm DE}$ and $\phi$ around the matter
point (b) is exactly the same as Eq.~(\ref{Omedephi}).

At the de Sitter point (c), both $H$ and $\phi$ are constants.
Provided that this fixed point is stable, the solutions with
different initial conditions finally converge to the
de Sitter attractor. {}From Eq.~(\ref{Omedephi}) we find
that, in the limit $p \gg 1$, the field $\phi$ stays nearly
constant during the radiation and matter eras, so the
cosmological dynamics should be close to that of
the $\Lambda$CDM model at the background model.
In fact, we have that $w_{\rm DE} \to -1$ in Eqs.~(\ref{casea})
and (\ref{caseb}) by taking the limit $p \to \infty$.

To discuss the stability of the fixed points (a), (b), (c),
we consider small perturbations
$\delta \Omega_{\rm DE}$ and $\delta \Omega_{r}$
around them. These homogeneous perturbations obey
\begin{eqnarray}
\left(
\begin{array}{c}
\delta \Omega_{\rm DE}' \\
\delta \Omega_{r}'
\end{array}
\right) = {\cal M}
\left(
\begin{array}{c}
\delta \Omega_{\rm DE} \\
\delta \Omega_{r}
\end{array}
\right)\,,\qquad
{\cal M}=\left( \begin{array}{cc}
\cfrac{\partial f_1}{\partial \Omega_{\rm DE}}&
\cfrac{\partial f_1}{\partial \Omega_{r}}\\
\cfrac{\partial f_2}{\partial \Omega_{\rm DE}}&
\cfrac{\partial f_2}{\partial \Omega_{r}}
\end{array} \right)\,,
\label{uvdif}
\end{eqnarray}
where $f_1$ and $f_2$ are the functions on the r.h.s. of
Eqs.~(\ref{dx1}) and (\ref{dx2}) respectively, and
the components of the matrix ${\cal M}$ should be
evaluated at the fixed points.
The eigenvalues of the $2 \times 2$ matrix ${\cal M}$
for the fixed points (a), (b), (c) are given,
respectively, by
\ba
& &{\rm (a)}~~\mu_1=4(1+s)\,,\quad \mu_2=1\,,\\
& &{\rm (b)}~~\mu_1=3(1+s)\,,\quad \mu_2=-1\,,\\
& &{\rm (c)}~~\mu_1=-3\,,
\quad \mu_2=-4\,.
\ea

If $s>-1$, then  the radiation fixed point (a) is unstable
due to the positivity of $\mu_1$ and $\mu_2$.
In this case the matter point (b) corresponds to
a saddle. The de Sitter solution (c) is always stable
because the two eigenvalues are negative.
Thus, the cosmological trajectory is characterized by
the sequence of the fixed points: (a)$\to$(b)$\to$(c).
The dark energy equation of state evolves
from $w_{\rm DE}=-1-4s/3$ (radiation era) to
$w_{\rm DE}=-1-s$ (matter era) and then it finally
approaches the value $w_{\rm DE}=-1$
at the de Sitter attractor.
The same cosmological evolution is realized for the tracker 
solution in extended scalar Galileon theories \cite{extended1}.
In this case the second time derivative 
$\ddot{\phi}$ is present in the background equations of motion, 
so there are initial conditions where the solutions 
are not on the tracker.
In generalized vector Galileon theories, the absence of the 
$\ddot{\phi}$ term in Eqs.~(\ref{be1})-(\ref{be3}) 
does not allow the deviation from the tracker solution 
explained above.

\begin{figure}
\begin{center}
\includegraphics[height=3.3in,width=3.3in]{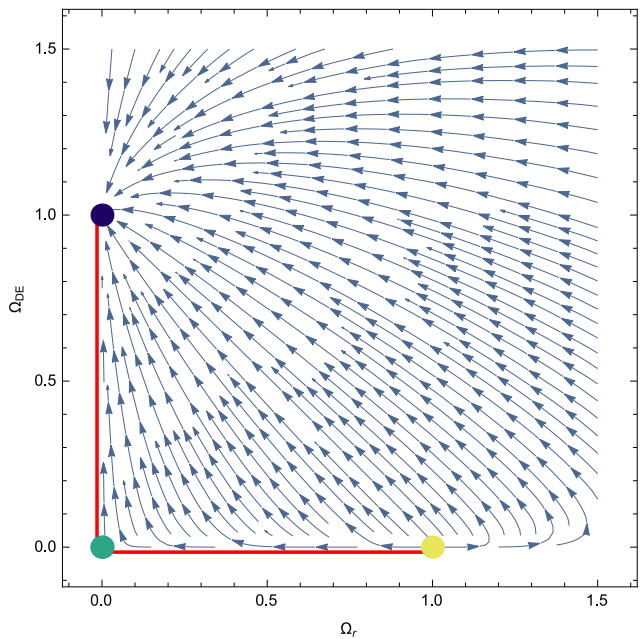}
\end{center}
\caption{\label{fig1}
This figure illustrates a specific example of the phase map portrait of
the dynamical autonomous system for $s=1$.
The radiation fixed point (a) is encoded by the yellow dot,
the matter fixed point (b) by green, and the
de Sitter fixed point (c) by blue.
The red line corresponds to the cosmological
trajectory explained in the main text. }
\end{figure}

In Fig.~\ref{fig1} we show an example of the phase-map portrait
of the dynamical autonomous system for $s=1$.
One can see that the de Sitter fixed point (c) is an attractor.
All the trajectories with arbitrary initial conditions end there.
The red line denotes the cosmological trajectory following the
sequence: (a)$\to$(b)$\to$(c).
We would like to remind that the reason why the zero component
of the vector field is not part of the dynamical autonomous system
comes from the fact that its equation of motion is an algebraic equation
that can be used to determine it fully by $H$.
This will be always the case as long as Eq.~(\ref{be3d}) does
not contain any time derivative applied on $\phi$.
Therefore, the dynamical system is reduced by one dimension.
This on the other hand will have as a consequence the known general
result that gravity with auxiliary fields gives rise to a modified matter coupling.

\begin{figure}
\begin{center}
\includegraphics[height=3.3in,width=3.5in]{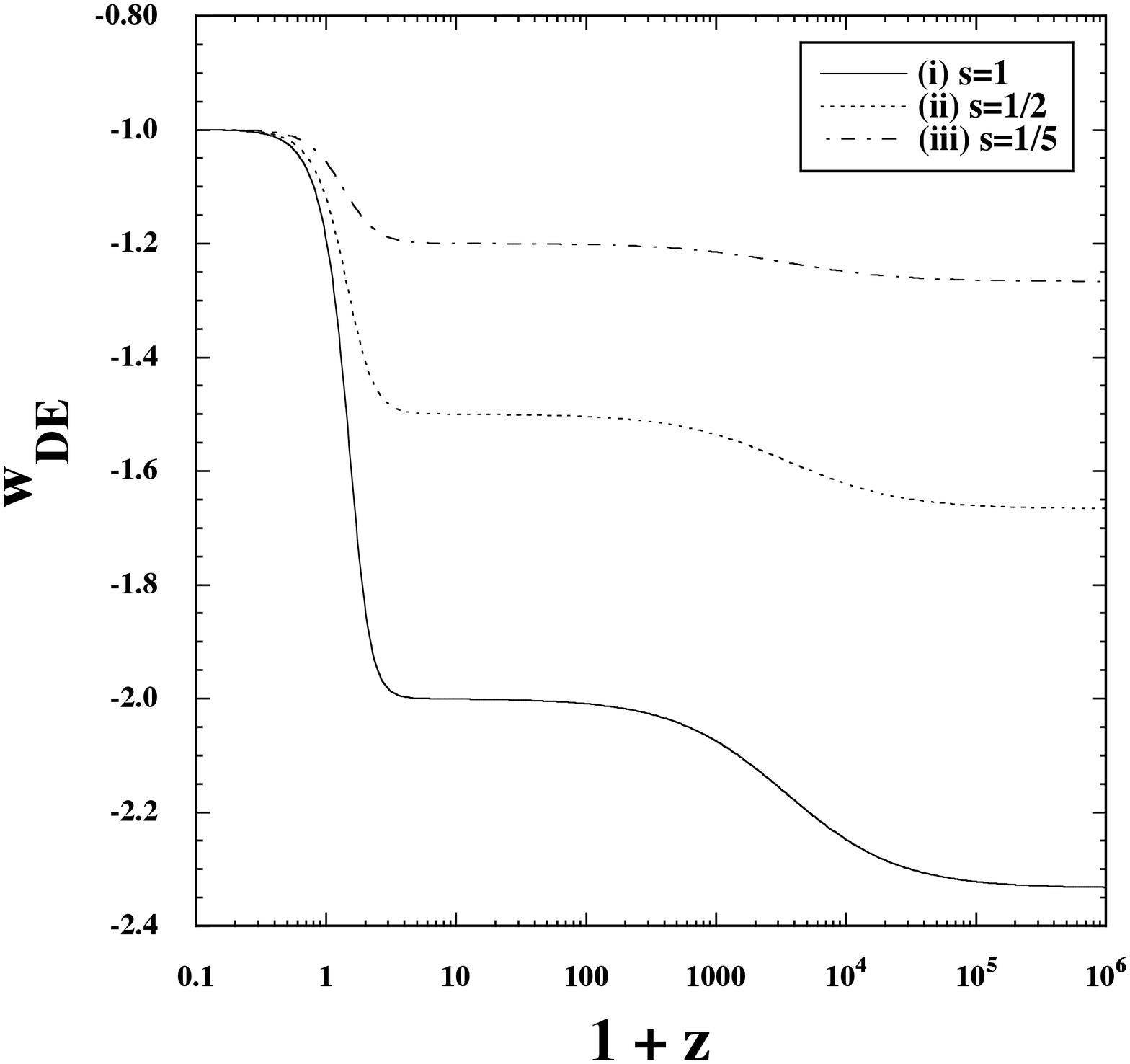}
\includegraphics[height=3.3in,width=3.2in]{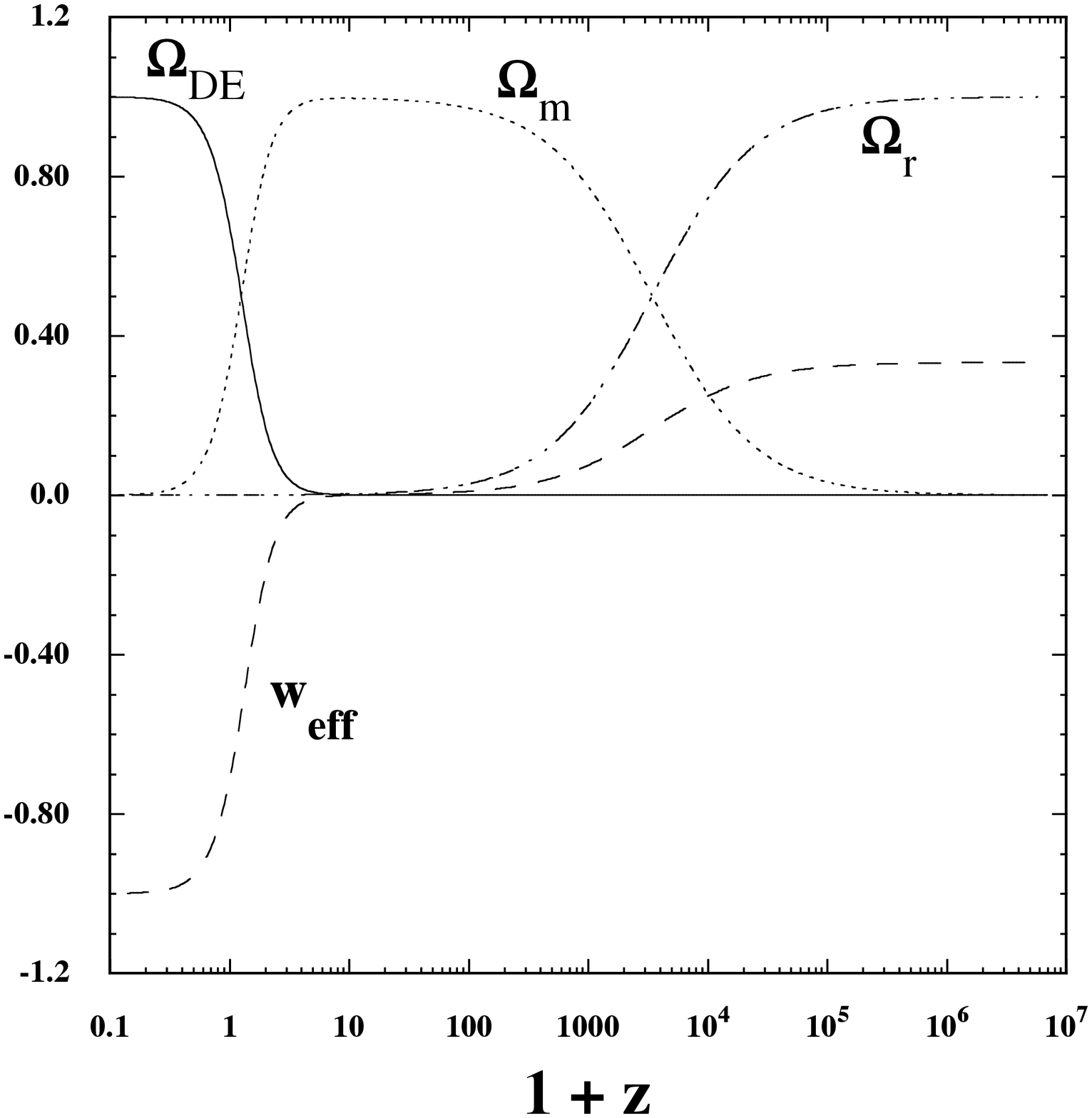}
\end{center}
\caption{\label{fig2}
(Left) Evolution of $w_{\rm DE}$ versus $1+z$ (where
$z$ is the redshift) for the three cases
(i) $s=1$, (ii) $s=1/2$, and (iii) $s=1/5$.
In each case the initial conditions are chosen to be
(i) $\Omega_{\rm DE}=1.0 \times 10^{-42}$,
$\Omega_{r}=0.998$ at
$z=1.7 \times 10^6$,
(ii) $\Omega_{\rm DE}=1.5 \times 10^{-36}$,
$\Omega_{r}=0.9996$ at
$z=9.0 \times 10^6$, and
(iii) $\Omega_{\rm DE}=7.0 \times 10^{-29}$,
$\Omega_{r}=0.9995$ at
$z=6.9 \times 10^6$.
(Right) Evolution of $w_{\rm eff}$, $\Omega_{\rm DE}$,
$\Omega_m$, and $\Omega_r$ for $s=1/5$
with the same initial conditions as those in the case (iii).}
\end{figure}

In the left panel of Fig.~\ref{fig2}, we plot the evolution of
$w_{\rm DE}$ versus $1+z$ (where $z\equiv a_0/a-1$ is the
redshift) for three different values of $s$.
We identify the present epoch ($z=0$) as
$\Omega_{\rm DE}=0.68$ and $\Omega_r=1 \times 10^{-4}$.
As estimated above, the dark energy equation of state for
$s=1$ (case (i) in Fig.~\ref{fig2}) evolves as
$w_{\rm DE}=-7/3$ (radiation era) $\to$
$w_{\rm DE}=-2$ (matter era) $\to$
$w_{\rm DE}=-1$ (de Sitter era).
This behavior of $w_{\rm DE}$ is the same as that of
the tracker solution found for scalar Galileons \cite{DT10}.
Unfortunately, this case is in tension with the combined
observational constraints of SNIa, CMB, and BAO
due to the large deviation of $w_{\rm DE}$ from $-1$
before the onset of cosmic acceleration \cite{Nesseris}.

For smaller $|s|$, the dark energy equation
of state in the radiation and matter eras tends to be closer
to the value $w_{\rm DE}=-1$. In the cases (ii) and (iii)
shown in the left panel of Fig.~\ref{fig2}, the values of
$w_{\rm DE}$ during the matter-dominated epoch are
given, respectively, by $w_{\rm DE}=-1.5$ and $w_{\rm DE}=-1.2$.
The likelihood analysis based on the SNIa, CMB,
and BAO data in extended scalar Galileon theories has given
the bound $0 \le s \le 0.36$ (95 \%\,CL) 
for tracker solutions \cite{extended2},
which can be also applied to the present model.
Hence the cases (i) and (ii) are in tension with the data,
but the case (iii) is observationally allowed.

In the right panel of Fig.~\ref{fig2}, we show the evolution
of $w_{\rm eff}$ and the density parameters
$\Omega_{\rm DE}$, $\Omega_m$, $\Omega_r$ for $s=1/5$.
As estimated analytically, the solution repels away from the unstable
radiation point (a) (with $\Omega_r=1$,
$w_{\rm eff}=1/3$) and temporally dwells in the
saddle matter point (b)
(with $\Omega_m=1$, $w_{\rm eff}=0$).
Finally, the solution approaches the stable de Sitter
attractor (c) characterized by
$\Omega_{\rm DE}=1$, $w_{\rm eff}=-1$.

\subsection{No-ghost and stability conditions}
\label{ghosec}

Let us study the no-ghost and stability conditions of
tensor, vector, and scalar perturbations for
the theories given by the functions (\ref{G2345})
with the powers (\ref{p345}).
In what follows, unless otherwise stated, we shall discuss 
the case in which $\beta_4$ and $\beta_5$ are 
non-zero\footnote{We note that under 
the limits $G_{4} \to M_{\rm pl}^2/2$ and $G_5 \to 0$ 
in Eqs.~(\ref{qT}), (\ref{cT}), (\ref{qv})
and (\ref{cv}), it follows that $q_T \to M_{\rm pl}^2$,
$c_T^2 \to 1$, $q_V \to 1$, and $c_V^2 \to 1$.}. 

{}From Eqs.~(\ref{qT}) and (\ref{cT}) we obtain
\ba
q_T &=& M_{\rm pl}^2 \left[ 1 -\Omega_{\rm DE}
\left\{ 1+\frac{1-2p_2\beta_5}{\beta}p_2(p+p_2)
\right\} \right]\,,\\
c_T^2 &=& 1+\frac{\Omega_{\rm DE}(p+p_2)\xi_T}
{(2p+2p_2-1)(p+p_2\Omega_{\rm DE})
[\beta(1-\Omega_{\rm DE})-p_2(p+p_2)
\Omega_{\rm DE}(1-2p_2\beta_5)]}\,,
\ea
where
\ba
\xi_T &\equiv& 2p\beta+p_2 \left[2pp_2 \left\{ 1-2(p_2-6)\beta_5
\right\}+9p_2\beta_5 (2p_2-1)+2p^2(1-2p_2\beta_5)\right]
+3\Omega_r p_2^2\beta_5(2p+2p_2-1) \nonumber \\
& & +\Omega_{\rm DE}p_2 \left[ 2\beta+
p_2(2p+2p_2+\beta_5 \{ 9-4p_2(3+p_2)
-2p(9+2p_2)\})\right]\,.
\ea
Taking the limit $\Omega_{\rm DE} \to 0$
in the radiation and matter dominated epochs,
we have $q_T \to M_{\rm pl}^2$ and $c_T^2 \to 1$, respectively.
At the de Sitter point (c) characterized by
$\Omega_{\rm DE}=1$ and $\Omega_r=0$,
the conditions for avoiding the tensor ghost and
instabilities are given by
\ba
\left( q_T \right)_{\rm dS} &=& -\frac{1-2p_2\beta_5}{\beta}p_2 (p+p_2)
M_{\rm pl}^2>0\,,
\label{qtdscon}\\
\left( c_T^2 \right)_{\rm dS} &=& \frac{1-12p_2\beta_4+4p_2\beta_5}
{1-2p_2 \beta_5}>0\,.
\label{ctdscon}
\ea
For the theories with $\beta_5=0$ and $p_2(p+p_2)>0$,
the conditions (\ref{qtdscon}) and (\ref{ctdscon})
translate to $\beta<0$ and $1-12p_2 \beta_4>0$, respectively.
When $\beta_5=0$, the condition for avoiding the tensor ghost
at the de Sitter point requires that $G_2<0$,
in which case the constant $b_2$ and
the quantity $y$ in Eq.~(\ref{ydef}) are negative.
This is consistent with the condition $\beta<0$
in the sense that the dark energy density parameter
(\ref{OmegaDE}) remains positive.
In this case the term $G_4$ appearing in Eq.~(\ref{ctds})
is given by $G_4=-M_{\rm pl}^2p_2(p+p_2)(1-12p_2\beta_4)/(2\beta)$,
which is positive under the two conditions
(\ref{qtdscon}) and (\ref{ctdscon}) (with $\beta_5=0$).

For vector perturbations, the quantities (\ref{qv}) and (\ref{cv})
reduce, respectively, to
\ba
q_{V} &=& 1-\frac{12\,\Omega_{\rm DE}\,p_2^2
(p+p_2)(c_2\beta_4+d_2\beta_5)}{u^2 \beta}\,,
\label{qVex}\\
c_{V}^2 &=& 1+\frac{3p_2^2(p+p_2)\Omega_{\rm DE}}{\beta q_V u^2}
\left[\frac{6p_2^2(p+p_2)(2\beta_4-\beta_5)^2\Omega_{\rm DE}}
{\beta q_T/M_{\rm pl}^2}
+\frac{\left\{(2p_2+3)\Omega_{\rm DE}-\Omega_r
+2p-3\right\}d_2\beta_5}
{p+p_2\Omega_{\rm DE}}\right]\,,
\label{cVex}
\ea
where
\be
u \equiv \frac{\phi}{M_{\rm pl}}\,.
\ee
The quantity $q_V$ contains the time-dependent
factors $\Omega_{\rm DE}$ and $u$.
{}From Eq.~(\ref{Omedephi}) it follows that
$\Omega_{\rm DE}$ grows faster than $u^2$ for
\be
p+p_2>1\,.
\label{pp2con}
\ee
Under this condition, the term containing
$\Omega_{\rm DE}$ in Eq.~(\ref{qVex}) tends
to be negligible relative to the $u^2 \beta$ term
as we go back to the past.
In other words, $q_V$ is close to 1 in the radiation
and deep matter eras.
At the de Sitter fixed point, the absence of the vector
ghost requires that
\be
\left( q_{V} \right)_{\rm dS} = 1-\frac{12\,p_2^2
(p+p_2)(c_2\beta_4+d_2\beta_5)}{u^2 \beta}>0\,.
\label{qVexds}
\ee

In the following we shall focus on the case in which
the powers $p$ and $p_2$ satisfy the condition
(\ref{pp2con}). Then, $c_V^2$
approaches 1 in the asymptotic past.
At the de Sitter point, the condition for avoiding the
instability of vector perturbations reads
\be
\left( c_{V}^2 \right)_{\rm dS} =
1-\frac{18p_2^3 (p+p_2)(2\beta_4-\beta_5)^2}{(1-2\beta_5p_2)
\beta\,u^2(q_V)_{\rm dS}}
+\frac{6d_2\beta_5p_2^2(p+p_2)}{\beta\,u^2(q_V)_{\rm dS} }>0\,.
\label{cVexds}
\ee
Under the no-ghost condition (\ref{qtdscon}) of the tensor mode,
the second term on the r.h.s. of Eq.~(\ref{cVexds})
is positive. If $d_2 \beta_5(p+p_2)/\beta \geq 0$ and
$(q_V)_{\rm dS}>0$, then the vector propagation 
speed is super-luminal.

For scalar perturbations, the quantity (\ref{qs}) reduces to
\be
q_S=2^{2-p_2}M_{\rm pl}^{2(1+p_2)}u^{2p_2}
p_2 \left( p+ p_2\Omega_{\rm DE} \right)\,
b_2 \left[ -1+6(2p+2p_2-1)\beta_4-2(3p+2p_2)\beta_5
\right]\,,
\label{qsex}
\ee
so that the scalar ghost is absent for
\be
p_2 \left( p+ p_2\Omega_{\rm DE} \right)\,
b_2 \left[ -1+6(2p+2p_2-1)\beta_4-2(3p+2p_2)\beta_5
\right]>0\,.
\ee
We need to caution that, even when $q_S>0$, there are cases
in which the term $w_1-2w_2$ appearing in the denominator of
$Q_S$ crosses 0. 
Due to the divergence of $Q_S$, we shall exclude such cases 
in the following discussion\footnote{The divergence of $Q_S$ does 
not necessarily imply theoretical inconsistencies. It may simply 
indicate an infinite weak coupling, depending on the nature of 
non-linear interactions. 
Nonetheless we do not consider this possibility just to be conservative.}. 
One can express $w_1-2w_2$ in the form
\be
w_1-2w_2=-2HM_{\rm pl}^2 \left( 1-\Omega_{\rm DE}w_c
\right)\,,\qquad
w_c \equiv 1+p+p_2-\frac{p_2(p+1)(p+p_2)(2\beta_5 p_2-1)}
{\beta}\,.
\ee
For $0<\Omega_{\rm DE}<1$, the term $w_1-2w_2$
remains negative for $w_c<1$, i.e.,
\be
\frac{p_2(p+1) (2\beta_5 p_2-1)}{\beta}>1\,.
\label{divcon}
\ee
To satisfy this inequality for the theories with $\beta_5=0$ and
$p_2(p+1)>0$, the necessary condition to be required is
$\beta<0$.

The general expression of $c_S$ computed from
Eq.~(\ref{cs}) is cumbersome, but we can obtain
its analytic expression in radiation and matter eras
by taking the limit $\Omega_{\rm DE} \to 0$.
This leads to the following condition
\ba
\left( c_S^2 \right)_{\rm early}
&=& \frac{3-5p-6p_2
+6\beta_4[10p^2+p(22p_2-27)+12p_2(p_2-2)+9]
-2\beta_5[9+3p(5p-11)+4p_2(3p_2+7p-6)]}
{6p^2[6\beta_4(2p+2p_2-1)-2\beta_5(3p+2p_2)-1]}
\nonumber \\
& &+\frac{\Omega_r[1-p-2p_2+6\beta_4(p+2p_2-3)
(2p+2p_2-1)-2\beta_5 \{3+3p(p-3)+4p_2(p_2+2p-2)\}]}
{6p^2[6\beta_4(2p+2p_2-1)-2\beta_5(3p+2p_2)-1]}>0\,.
\label{csex1}
\ea
At the de Sitter fixed point, the scalar instability
is absent for
\ba
\left( c_S^2 \right)_{\rm dS}=\frac{\eta\left[(p+p_2)\eta
-\beta\left\{\beta+p_2(1+p)(1-2p_2 \beta_5)\right\} 
(q_V)_{\rm dS}\,u^2\right]}
{6 p_2 \beta^2 (2 p_2 \beta_5-1)
\left[\beta+p p_2 (1-2 p_2 \beta_5)\right] 
(q_V)_{\rm dS}\,u^2}>0\,,
\label{csdeex}
\ea
where
\ba
\eta&\equiv&
p_2^2 \left[ \beta+p_2 (1+p) (1-2 p_2 \beta_5)\right]
\left[ 1+6 (5-2p-2p_2) \beta_4 -2 (6-3p-2p_2) \beta_5\right]\notag\\
&& +\left[\beta+pp_2 (1-2p_2 \beta_5)\right]
\left[\beta+p_2 (p-1) (1-2p_2\beta_5)\right]\,.
\ea

For the computation of the dimensionless field $u=\phi/M_{\rm pl}$
appearing in $q_V, c_V^2, q_S, c_S^2$,
we introduce the following dimensionless quantity
\be
\lambda \equiv \left( \frac{\phi}{M_{\rm pl}} \right)^{p}
\frac{H}{m}\,,
\ee
where $\lambda$ is constant and the mass $m$ is related to
the coefficient $b_2$ in $G_2$, as
\be
G_2=-m^2M_{\rm pl}^{2(1-p_2)}X^{p_2}\,.
\ee
On using Eq.~(\ref{OmegaDE}) with the definition of $y$
given in Eq.~(\ref{ydef}),
we can express $u$ in the form
\be
u=\left[ -2^{p_2} \frac{3\lambda^2p_2(p+p_2)\Omega_{\rm DE}}{\beta}
\right]^{\frac{1}{2(p+p_2)}}\,.
\ee
We are interested in the case where the today's values of $\phi$
and $H$ are of the orders of $M_{\rm pl}$ and $m$,
respectively, so that $\lambda={\cal O}(1)$.
Such a very light mass of the vector field is only relevant
to the physics associated with the late-time acceleration.
In local regions of the Universe the effect of the mass
term is negligible, but the derivative interactions like
$b_4X^2$ in $G_4$ play crucial roles to screen
the propagation of the fifth force mediated by the
vector field \cite{Tasinato,scvector}.

\subsection{Models with $p_2=1$ and $G_5=0$}
\label{G50sec}

To be more concrete, we shall first study the models with
$G_5=0$, $p_2=1$, and $p>0$.
In this case, the parameter $\beta$ defined
in Eq.~(\ref{betadef}) reduces to
\be
\beta=6(2p+1)\beta_4-p-1\,.
\ee
To avoid the divergence of $Q_S$ associated with the
sign change of $w_1-2w_2$, we require the condition
(\ref{divcon}). In the present case, this translates to
\be
0<\beta_4<\frac{p+1}{6(2p+1)}\,.
\label{be4con}
\ee
The upper bound of $\beta_4$ comes from the condition $\beta<0$.

\begin{figure}
\begin{center}
\includegraphics[height=3.3in,width=3.3in]{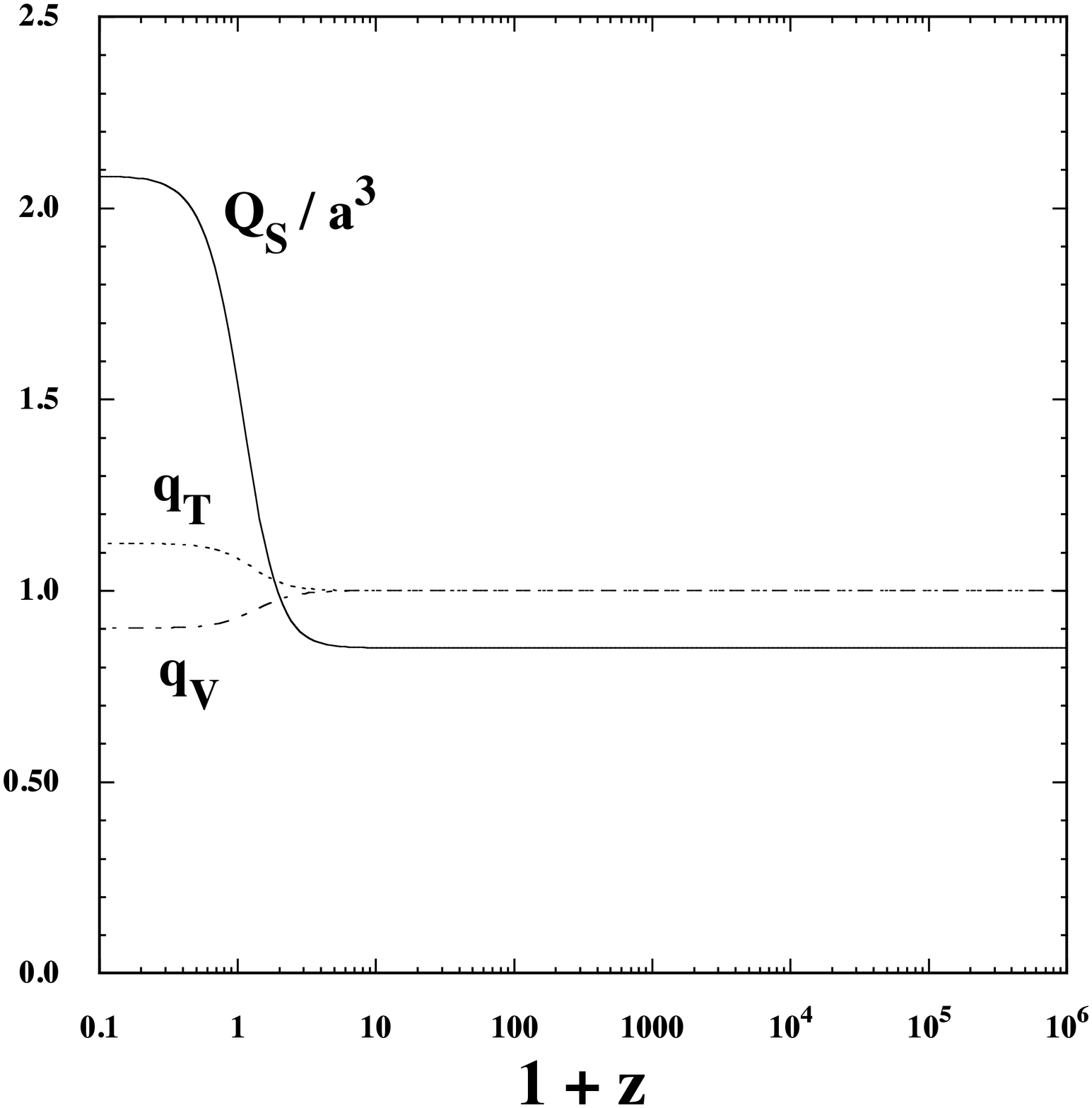}
\includegraphics[height=3.3in,width=3.3in]{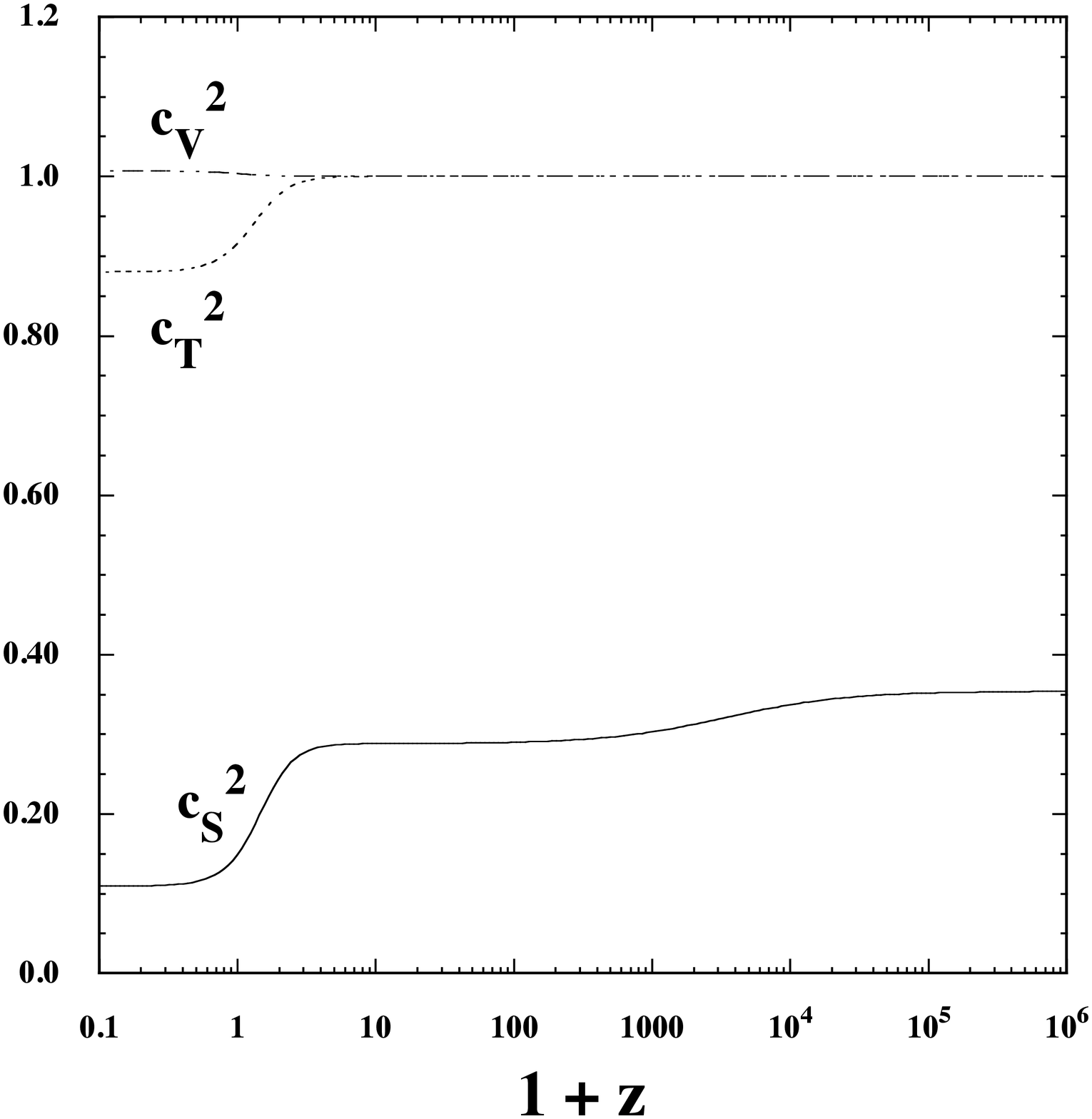}
\end{center}
\caption{\label{fig3}
Evolution of $q_T, q_V, Q_S/a^3$ (left) and
$c_T^2, c_V^2, c_S^2$ (right) versus $1+z$
for $p_2=1$, $p=5$, $\beta_4=0.01$, $\beta_5=0$,
$c_2=-1$, and $\lambda=1$.
Note that $q_T$ and $Q_S/a^3$ are normalized by
$M_{\rm pl}^2$ and $m^2$, respectively.
The background initial conditions are chosen to be
$\Omega_{\rm DE}=7.0 \times 10^{-29}$
and  $\Omega_{r}=0.9995$ at $z=6.9 \times 10^6$. }
\end{figure}

The conditions (\ref{qtdscon}) and (\ref{ctdscon}) for tensor
perturbations at the de Sitter point reduce, respectively, to
\ba
\left( q_T \right)_{\rm dS} &=&
M_{\rm pl}^2 \left[ 1-\frac{6(2p+1)\beta_4}{p+1}
\right]^{-1}>0\,,
\label{qtten}\\
\left( c_T^2 \right)_{\rm dS} &=& 1-12\beta_4>0\,.
\label{ctten}
\ea
The condition (\ref{qtten}) is automatically satisfied
under the inequality (\ref{be4con}).
On the other hand the condition (\ref{ctten}) gives the
upper bound on $\beta_4$ tighter  than Eq.~(\ref{be4con}),
so the allowed range of $\beta_4$ shrinks to
\be
0<\beta_4<\frac{1}{12}\,.
\label{becon2}
\ee
In Fig.~\ref{fig3} we plot the evolution of $q_T$ and
$c_T^2$ for $\beta_4=0.01$ and $p=5$.
As estimated in Sec.~\ref{ghosec}, we have
$q_T \simeq M_{\rm pl}^2$ and $c_T^2 \simeq 1$
during the radiation and matter eras.
Finally, we see that $q_T$ and $c_T^2$ approach the asymptotic
values (\ref{qtten}) and (\ref{ctten}), respectively.
Under the condition (\ref{becon2}), $(c_{T}^2)_{\rm dS}$ is
smaller than 1. If the today's value of $c_T^2$ is smaller than 1,
the tensor propagation speed squared is constrained to be
$1-c_T<2 \times 10^{-15}$ from the gravitational Cherenkov
radiation \cite{Cherenkov}.
On using the value (\ref{ctten}) at the de Sitter fixed point,
we obtain the bound $\beta_4 < {\cal O}(10^{-16})$.
In this case, we have numerically confirmed that the
Cherenkov-radiation constraint of $c_T$ is satisfied today.

For vector perturbations, the conditions (\ref{qVexds})
and (\ref{cVexds}) can be expressed, respectively, as
\ba
\left( q_V \right)_{\rm dS}
&=& 1-\frac{2c_2(p+1)(\beta+p+1)}
{(2p+1)\beta u^2}>0\,,\\
\left( c_V^2 \right)_{\rm dS} &=&
1+\frac{2(p+1)(\beta+p+1)^2}
{(2p+1)[2c_2(p+1)(\beta+p+1)-(2p+1)\beta u^2]}>0\,.
\ea
Since $\beta+p+1=6(2p+1)\beta_4$, we have
that $\left( q_V \right)_{\rm dS} \to  1$ and
$\left( c_V^2 \right)_{\rm dS} \to 1$ in the limit
$\beta_4 \to 0$. As we see in Fig.~\ref{fig3}, both
$q_V$ and $c_V^2$ are very close to 1 in the early
cosmological epoch. At late times $q_V$ and $c_V^2$
start to deviate from 1, but both of them are still
close to 1 for $\beta_4 \ll 1$ and $|c_2| \lesssim {\cal O}(1)$.

The no-ghost condition for scalar perturbations corresponds to
\be
q_S=2m^2M_{\rm pl}^4 \left(p+\Omega_{\rm DE} \right)
\left[ 1-6(2p+1)\beta_4 \right]u^2>0\,,
\label{qsexa}
\ee
which is satisfied for
\be
\beta_4<\frac{1}{6(2p+1)}\,.
\label{becon3}
\ee
If $p>1/2$, then the condition (\ref{becon3}) gives a tighter upper
bound on $\beta_4$ than that constrained from Eq.~(\ref{becon2}).

{}From Eq.~(\ref{csex1}) the conditions for avoiding the
Laplacian instability during the radiation and matter eras read
\ba
\left( c_S^2 \right)_{r}
&=& \frac{3p+2-6(6p^2-3p-2)\beta_4}
{3p^2[1-6(2p+1)\beta_4]}>0\,,\label{csr}\\
\left( c_S^2 \right)_{m}
&=& \frac{5p+3-6(10p^2-5p-3)\beta_4}
{6p^2[1-6(2p+1)\beta_4]}>0\,,\label{csm}
\ea
respectively.
The condition (\ref{csdeex}) at the de Sitter point reduces to
\be
\left( c_S^2 \right)_{\rm dS}
=\frac{2[1+6\beta_4(12\beta_4-1)(2p+1)]
[1+p+3\beta_4(2p+1)\{2(p+1) [ 6\beta_4(c_2+2)-1]-u^2\beta\}]}
{3\beta[1- 6(2p+1)\beta_4][u^2\beta-12c_2(p+1)\beta_4]}
>0\,.
\label{csdef}
\ee
In the limit that $\beta_4 \to 0$, we have the following
asymptotic values
\be
\left( c_S^2 \right)_{r} \to \frac{3p+2}{3p^2}\,,\qquad
\left( c_S^2 \right)_{m} \to \frac{5p+3}{6p^2}\,,\qquad
\left( c_S^2 \right)_{\rm dS} \to \frac{2}{3(p+1)u^2}\,,
\label{cSesti}
\ee
which are positive.
Hence the stability of scalar perturbations is always
ensured for $\beta_4$ close to 0.

The evolution of $Q_S/a^3$ and $c_S^2$ plotted
in Fig.~\ref{fig3} corresponds to the model parameters
$\beta_4=0.01$ and $p=5$.
In this case, the constant $\beta_4$ is in the range
consistent with both Eqs.~(\ref{becon2}) and (\ref{becon3}).
In the radiation and matter eras the parameter $q_S$, which
is positive, grows as $q_S \propto u^2 \propto t^{2/p}$.
Since both $q_S/\phi^2$ and $H^2/(w_1-2w_2)^2$ are constants
in this regime, $Q_S/a^3$ stays constant for $p_2=1$.
The quantity $q_S$ asymptotically approaches the value (\ref{qsexa})
at the de Sitter point characterized by $\Omega_{\rm DE}=1$ and constant $u$.
Provided that $\lambda={\cal O}(1)$, the dimensionless field $u$ is of
the order of 1 at the de Sitter attractor, so the mass
scale $m$ is of the order of the today's Hubble parameter $H_0$.

The analytic estimations of $q_S$ as well as
$Q_S/a^3$ show good agreement with our numerical results.
We also confirm that the term $w_1-2w_2$ always remains
negative in the numerical analysis of Fig.~\ref{fig3}.
If we choose the values of $\beta_4$ which are outside the range of
Eqs.~(\ref{be4con}) (e.g., negative $\beta_4$), we find that
$Q_S/a^3$ exhibits a divergence due to the sign change of
$w_1-2w_2$ around the transient epoch to the cosmic acceleration.
For the cosmological viability of the model with $\beta_5=0$ and $p_2=1$,
we require the conditions (\ref{becon2}) and (\ref{becon3}).

For $\beta_4=0.01$ and $p=5$ the analytic estimations (\ref{csr}) and
(\ref{csm}) give $\left( c_S^2 \right)_{r} \simeq  0.354$
and $\left( c_S^2 \right)_{m} \simeq 0.288$, respectively,
which are in good agreement with the numerical values
of $c_S^2$ in Fig.~\ref{fig3}.
Substituting the numerical value $u \simeq 1.172$ at the
de Sitter point into Eq.~(\ref{csdef}) with $c_2=-1$, we obtain
$(c_S^2)_{\rm dS}=0.109$.
In Fig.~\ref{fig3}, we see that $c_S^2$ is always positive
during the cosmic expansion history, so there is no
Laplacian instability of scalar perturbations.

While the numerical evolution of Fig.~\ref{fig3} corresponds to
the case ($\beta_4=0.01$) in which the Cherenkov-radiation constraint of $c_T^2$
is not imposed, we have also obtained the numerical evolution of
$q_T,q_V,q_S$ and $c_T^2,c_V^2,c_S^2$ under the bound
$0<\beta_4 \lesssim 10^{-16}$. In such cases the three no-ghost
conditions are trivially satisfied, with $c_T^2$ and $c_V^2$
very close to 1. In addition, $c_S^2$ is close to the positive values
given by Eqs.~(\ref{cSesti}), so there is no instability of scalar perturbations.

\subsection{Models with general $p_2$ and $G_5$ realizing $c_T^2>1$}

In the models with $p_2=1$ and $G_5=0$, the tensor propagation
speed is sub-luminal under the condition (\ref{becon2}).
In this case, the constraint from Cherenkov radiation puts
a tight bound on $\beta_4$.
Let us study the possibility for realizing $c_T^2>1$ to escape the
Cherenkov-radiation constraint for general $p_2$ and $G_5$,
while avoiding the divergence of $Q_S$.
Provided that
\be
p_2\beta_5<1/2\,,
\label{p2beta5}
\ee
the tensor propagation speed squared (\ref{ctdscon}) at the
de Sitter solution is larger than
or equal to 1 for
\be
\beta_5\geq 2\beta_4\,. 
\label{be5con1}
\ee
Under Eq.~(\ref{p2beta5}) with negative $\beta$,
the condition (\ref{divcon}) for avoiding the divergence of $Q_S$ reads
\be
\frac{(p+p_2)(4p_2\beta_5+1)+(p+1)(2p_2\beta_5-1)}
{6p_2(2p+2p_2-1)}<\beta_4<\frac{(p+p_2)(4p_2\beta_5+1)}
{6p_2(2p+2p_2-1)}\,,
\label{be5con2}
\ee
where we have assumed $p_2(2p+2p_2-1)>0$.

When $p_2=1$ the first inequality of (\ref{be5con2}) reads
$\beta_5<\beta_4(2p+1)/(p+1)$, so this is not compatible
with the condition (\ref{be5con1}) for $p>0$.
Thus, for $p_2=1$, the sub-luminal property of $c_T^2$ found for
$G_5=0$ in Sec.~\ref{G50sec} also holds for the theories
with $G_5 \neq 0$ and $p>0$.
For $p_2$ different from 1, it is possible to simultaneously
fulfill the conditions (\ref{p2beta5})-(\ref{be5con2}).
If $p_2<1$, for example, the simple case with 
$\beta_4=\beta_5=0$ satisfies such three conditions. 
The allowed parameter space also exists for non-zero values 
of $\beta_4$ and $\beta_5$.

\begin{figure}
\begin{center}
\includegraphics[height=3.3in,width=3.3in]{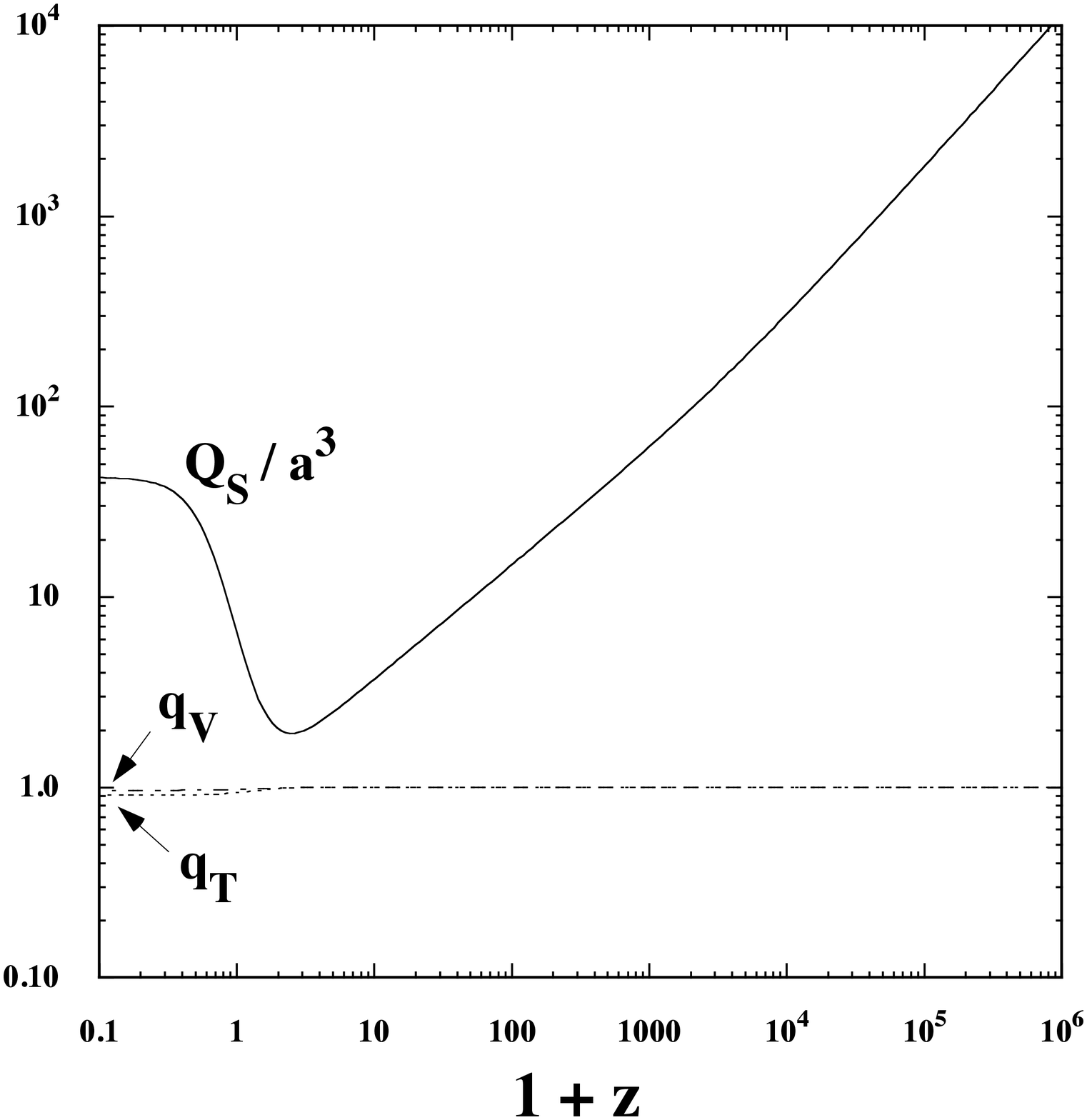}
\includegraphics[height=3.3in,width=3.3in]{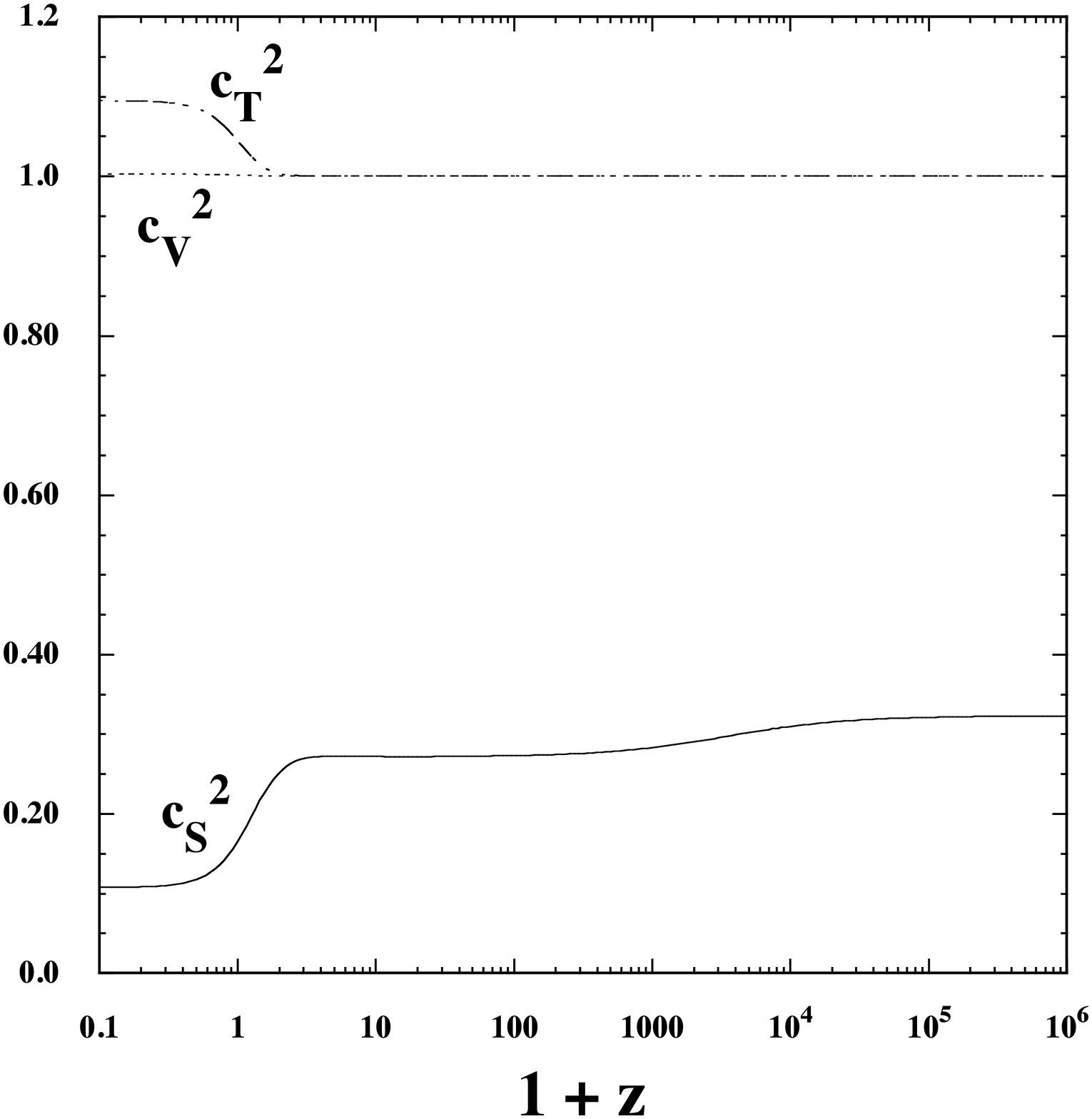}
\end{center}
\caption{\label{fig4}
Evolution of $q_T, q_V, Q_S/a^3$
(left) and $c_T^2, c_V^2, c_S^2$ (right) versus $1+z$
for $p_2=1/2$, $p=5/2$, $\beta_4=0.01$, $\beta_5=0.05$,
$c_2=-1$, $d_2=0$, and $\lambda=1$
with $G_2=-m^2M_{\rm pl}X^{1/2}$.
The quantities $q_T$ and $Q_S/a^3$ are normalized by
$M_{\rm pl}^2$ and $m^2$, respectively.
The background initial conditions are chosen to be
the same as those in Fig.~\ref{fig3}. }
\end{figure}

In Fig.~\ref{fig4} we plot the evolution of $q_T,q_V,Q_S/a^3$ and
$c_T^2,c_V^2,c_S^2$ for $p_2=1/2$, $p=5/2$,
$\beta_4=0.01$, and $\beta_5=0.05$, under which  the conditions
(\ref{p2beta5})-(\ref{be5con2}) are satisfied.
The value of $\lambda=(\phi/M_{\rm pl})^p(H/m)$ is chosen to
be $1$ with the negative function $G_2=-m^2M_{\rm pl}X^{1/2}$.
In the left panel of Fig.~\ref{fig4}  we find that the no-ghost conditions
are satisfied during the cosmic expansion history.
During the radiation and matter eras the quantity $q_S$ evolves as
$q_S \propto \phi^{2p_2}$, so that $Q_S/a^3$ decreases in
proportion to $\phi^{-1}$. After the end of the matter era,
$Q_S/a^3$ approaches the value at the de Sitter point
without any divergence.
In Fig.~\ref{fig4} we find that $c_T^2$ is larger than 1 today,
so the bound of Cherenkov radiation is not applied to this case.
Since both $c_V^2$ and $c_S^2$ are positive for the model
parameters used in Fig.~\ref{fig4},
there are no Laplacian instabilities of vector and scalar
perturbations as well.

The case shown in Fig.~\ref{fig4} is one example for realizing
$c_T^2>1$ without the divergence of $Q_S$.
There are also other cases in which
ghosts and Laplacian instabilities are absent with $c_T^2>1$.
We leave the detailed analysis about the viable parameter space
consistent with observations and experiments for a future work.

\section{Conclusions}
\label{consec}

We have studied the cosmology in generalized Proca theories 
with three propagating degrees of freedom
(besides two tensor gravitational degrees of freedom) 
in the presence of a matter fluid.
The temporal component $\phi$ of the vector field can drive
the cosmic acceleration due to the existence of its derivative interactions.
By construction of the action (\ref{Lag}), there is a branch of the background solution
where the field $\phi$ depends on the Hubble expansion rate $H$ alone. 
This allows the existence of de Sitter solutions characterized by 
constant $\phi$ and $H$.

Expanding the action  (\ref{Lag}) up to second order in cosmological
perturbations on the flat FLRW background, we have derived
general conditions for the absence of ghosts and Laplacian instabilities
in the small-scale limit. Unlike scalar-tensor theories, the propagation of
vector modes is non-trivial in the sense that the intrinsic vector Lagrangians
proportional to the constants $c_2$ and $d_2$ affect the no-ghost condition
and the propagation speed of vector perturbations.
In total, there are six no-ghost and stability conditions associated
with tensor, vector, and scalar perturbations.

Applying our general results to de Sitter solutions, we find that, in the limit
$G_5 \to 0$, the function $G_2$ needs to be negative to avoid
the tensor ghost, but this does not necessarily imply the appearance
of tachyons. In fact, defining vector perturbations with the background
tetrad basis, the mass squared of the vector mode is positive
on the de Sitter solution.
In the early cosmological epoch the tensor propagation speed squared
is close to 1, but this is not generally the case for scalar and vector
perturbations.

We have also proposed a class of concrete models
in which the field $\phi$ grows as $\phi^{p} \propto H^{-1}$ ($p>0$) 
with the decrease of $H$.
The functions $G_{2,3,4,5}$ realizing this solution is given by
Eq.~(\ref{G2345}) with the powers satisfying
(\ref{p345}), which accommodate vector
Galileons as a specific case ($p_2=1$ and $p=1$).
We have shown the existence of a de Sitter fixed point which is
always stable. Before reaching the de Sitter attractor,
the dark energy equation of state evolves as $w_{\rm DE}=-1-4s/3$
(radiation era) $\to$ $w_{\rm DE}=-1-s$ (matter era), where
$s=p_2/p$. For the consistency with the observations of SNIa, CMB, and
BAO, the parameter $s$ is constrained to be $0<s \le 0.36$.

For the concrete model proposed in Sec.~\ref{covasec}, we have
discussed the six no-ghost and stability conditions to search for
the theoretically allowed parameter space.
For avoiding the divergent behavior of $Q_S$
during the transient epoch to the cosmic acceleration,
there is an additional condition to be imposed.
For the theories with $G_5=0$ and $p_2=1$,
the parameter $\beta_4$ is constrained as
Eqs.~(\ref{becon2}) and (\ref{becon3}),
in which case the tensor propagation speed $c_T$ is sub-luminal.
On using the experimental constraint of $c_T$ from
the Cherenkov radiation, the upper bound of $\beta_4$ is
tightly constrained ($\beta_4 \lesssim 10^{-16}$).
For more general theories characterized by $G_5 \neq 0$ and/or $p_2 \neq 1$,
we find that there exists the parameter space in which
all the theoretically consistent conditions are satisfied
with $c_T^2\geq 1$. 

We have thus shown that the generalized Proca theories can
give rise to the dark energy dynamics compatible with theoretically
consistent conditions.
It will be of interest to place observational and experimental
constraints on the parameter space further by taking into
account the data of the growth rate of matter perturbations
as well as the bound of the tensor propagation speed.

Finally, we would like to point out that there might be interesting consequences 
in the presence of the sixth-order interactions 
$\mathcal{L}_6= G_6(X)L^{\mu\nu\alpha\beta} 
\nabla_\mu A_\nu \nabla_\alpha A_\beta + G_{6,X}(X) \tilde{F}^{\mu\nu}
\tilde{F}^{\alpha\beta} \nabla_\alpha A_\mu \nabla_\beta A_\nu/2$ \cite{Peter,Jimenez:2016isa}, 
where $L^{\mu\nu\alpha\beta}$ is the double dual Riemann tensor and 
$\tilde{F}^{\mu\nu}$ is the dual of the strength tensor. 
Since the background vector field considered here has only the 
zero component non-vanishing, these sixth-order interactions do not 
contribute at the background level, and similarly the tensor perturbations 
are not modified by them either, but they do have non-trivial effects 
on the vector and scalar perturbations. 
The implications of these sixth-order interactions will be investigated in a future work.

\section*{Acknowledgements}
We would like to thank J.~Beltran Jimenez, R.~Brandenberger, M.~Motta, 
J.~Yokoyama, and G.~Zhao for very useful discussions. 
LH acknowledges financial support from Dr.~Max R\"ossler, 
the Walter Haefner Foundation and the ETH Zurich Foundation.
RK is supported by the Grant-in-Aid for Research Activity Start-up
of the JSPS No.\,15H06635.
The work of SM is supported in part by Grant-in-Aid for Scientific Research 24540256 
and World Premier International Research Center Initiative (WPI), 
Ministry of Education, Culture, Sports, Science and Technology (MEXT), Japan.
ST is supported by the Grant-in-Aid for Scientific Research Fund 
of the JSPS No.\,24540286,
MEXT KAKENHI Grant-in-Aid for Scientific Research
on Innovative Areas ``Cosmic Acceleration'' (No.\,15H05890),
and the cooperation program between Tokyo
University of Science and CSIC. 
YZ is supported by the Strategic Priority Research Program
``The Emergence of Cosmological Structures" of the Chinese
Academy of Sciences, Grant No. XDB09000000.


\end{document}